\documentclass[12pt]{article}
\usepackage{amsmath,amssymb}
\usepackage{graphicx}
\usepackage{caption}
\usepackage{subcaption}
\usepackage{setspace,braket}

\def\be{\begin{equation}}
\def\ee{\end{equation}}
\def\lp{\ell_P}

\begin{document}
\title{ \begin{flushright}
\small IPMU14-0119 \vspace{1 cm} 
\end{flushright}
\bf Nonlinear constraints on gravity from entanglement}
\author{Shamik Banerjee$^{1,a}$, Apratim Kaviraj$^{2,b}$ and Aninda Sinha$^{2,c}$ 
\\
\it$^{1}$  Kavli Institute for the Physics and Mathematics of the Universe (WPI),\\ \it Todai Institutes for Advanced Study, \\ \it The University of Tokyo, Kashiwa, Chiba 277-8583, Japan\\
\it $^{2}$ Centre for High Energy Physics,
\it Indian Institute of Science,\\ \it C.V. Raman Avenue, Bangalore 560012, India. \\
{\small email: $^a$banerjeeshamik.phy@gmail.com; $^b$apratim, $^c$asinha@cts.iisc.ernet.in} \\}
\maketitle
\vskip 1cm
\begin{abstract}{\small 
Using the positivity of relative entropy arising from the Ryu-Takayanagi
formula for spherical entangling surfaces, we obtain constraints at the
nonlinear level for the gravitational dual. We calculate the Green's
function necessary to compute the first order correction to the entangling
surface and use this to find the relative entropy for non-constant stress
tensors in a derivative expansion. We show that the Einstein value
satisfies the positivity condition while the multi-dimensional parameter
space away from it gets constrained.}
\end{abstract}
\tableofcontents
\onehalfspace
\section{Introduction}

Can one derive Einstein field equations from quantum information? This question makes sense in the AdS/CFT correspondence. The precise version is as follows. The Ryu-Takayanagi entropy functional \cite{ryu} gives us a way to compute entanglement entropy in a conformal field theory dual to Einstein gravity. This entropy functional can be derived by assuming that the bulk theory was Einstein gravity \cite{maldacena}. We can turn the question around \cite{einstein1, einstein2, einstein4}. Given the form of the holographic entropy functional (eg. Ryu-Takayanagi), what are the bulk equations of motion?

In order to make progress, we appeal to a consistency condition that entanglement entropy has to satisfy: relative entropy has to be positive \cite{myers}. This condition is an inequality and involves comparing two density matrices. On the bulk side, this corresponds to adding perturbations to the AdS metric. At linearized order, the inequality is saturated and recently it has been possible to show that one recovers linearized gravitational equations of motion, not only for the Ryu-Takayanagi entropy functional but for any entropy functional that corresponds to higher derivative gravity in the bulk \cite{einstein3}. Since the gravity equations of motion are nonlinear it becomes an interesting question to ask what happens at nonlinear order. For the Ryu-Takayanagi entropy functional, there could be two possibilities: (a) we recover precisely Einstein equations or (b) we get a wider class of theories than just Einstein theory. Since the tool at hand is an inequality at first sight (a) appears to be a distant possibility. We can pose possibility (b) in a more precise way. We will add perturbations to the AdS metric. At linear order, we know that we recover linearized Einstein equations. At next order, we will write down the most general possible terms with undetermined coefficients. We will ask if the values that these coefficients can take are bounded. 

Progress in this direction was made in \cite{constrgrav}. Let us briefly review our findings there. 
We start by writing $d+1$ dimensional AdS space as \cite{skenderis}
\be\label{fgmetric}
ds^2=\frac{L^2}{z^2}dz^2+g_{\mu\nu}dx^\mu dx^\nu\,.
\ee
\be\label{metricpert}
g_{\mu\nu}=\frac{L^2}{z^2}\left[\eta_{\mu\nu}+a z^d T_{\mu\nu} +a^2 z^{2d}(n_1T_{\mu\alpha}T_\nu^\alpha+n_2\ \eta_{\mu\nu}T_{\alpha\beta}T^{\alpha\beta})+\cdots\right]\,,
\ee
where (\ref{metricpert}) is the most general term one can write down at quadratic order in the perturbation $T_{\mu\nu}$ for a constant stress tensor. The metric has a dimensionless constant $a=\frac{2}{d}(\ell_P/L)^{d-1}$, that can simply be absorbed in the $T_{\mu\nu}$. So in the rest of the paper we will not carry the constant anymore. The Ryu-Takayanagi entropy functional is given by
\be\label{RT}
S=\frac{2\pi}{\lp^{d-1}}\int d^{d-1}x\sqrt{h}\,,
\ee
where
\be
h_{ij}=\frac{L^2}{z^2}\left( g_{ij}+\partial_i z \partial_j z \right)\,,
\ee
where $z=z(x_i)$ is the bulk co-dimension two entangling surface (we will only consider static situations).
We will consider a spherical entangling surface for which the modular hamiltonian and hence the expression for relative entropy is known \cite{myers}.  The positivity of relative entropy leads to $\Delta H\geq \Delta S$ where $\Delta H$ is the difference in the expectation value of the modular hamiltonian with respect to the two density matrices and $\Delta S$ is the difference in the von Neumann entropies of the two density matrices--we will calculate this using eq.(\ref{RT}) in holography. Since  the change in the modular hamiltonian depends linearly on the stress tensor \cite{myers}, at quadratic order in the stress tensor $\Delta^{(2)}S$ must be negative.  Thus when we talk about nonlinear constraints on the metric, we have to compute $\Delta^{(2)}S$ and demand that it is negative--this will constrain the nonlinear terms in the metric. The solution for the entangling surface is given by $z_0^2=R^2+x_i x_i$. The perturbed surface can be written\footnote{$\epsilon$ is a small parameter just to keep track of the order of the perturbation.} as $z=z_0+\epsilon z_1$ where $z_1$ satisfies
\be \label{gf}
\frac{1}{z_0{}^{d-1}R}\left(\partial ^2\left(z_0z_1\right)-\frac{x^ix^j}{R^2}\partial _i\partial _j\left(z_0z_1\right)\right)=
\frac{z_0}{2R}\left(T\left(d-2\right)+T_x\left(d+2\right)\right)
\ee
The solution was  guessed in \cite{myers},
\be\label{z1constT}
z_1=-\frac{R^2z_0^{d-1}}{2(d+1)}(T+T_x)\,.
\ee
Using the above solution we get the second order correction to the difference in the entanglement entropy to be (see \cite{constrgrav}),
\be\label{del2SconstT}
\Delta^{(2)}S=2\pi (L/\ell_P)^{d-1}\Omega_{d-2}\left(
C_1T^2+C_2T_{ij}^2+C_3T_{i0}^2 \right)\,,
\ee
with
\begin{align}
\begin{split}
\label{correctC}
C_1&=\frac{2^{-3-d}d\left(1+4\left(d^2-1\right)n_2\right)\sqrt{\pi }R^{2d}\Gamma [d+1]}{\left(d^2-1\right)\Gamma \left[\frac{3}{2}+d\right]}\,,\\
C_2&=\frac{2^{-3-d} d  \sqrt{\pi } R^{2 d} \Gamma[1+d]}{\left(d^2-1\right) \Gamma\left[\frac{3}{2}+d\right]}\left(-1-2 d+4 (d+1) n_1 +4 \left(d^2-1\right) n_2\right)\,,\\
C_3&=-\frac{2^{-1-d} d (n_1+2 (d-1) n_2) \sqrt{\pi } R^{2 d} \Gamma[1+d]}{(d-1) \Gamma\left[\frac{3}{2}+d\right]}\, .
\end{split}
\end{align}

Now we must demand that $\Delta^{(2)}S\leq 0$. We can write $\Delta^{(2)}S= V^T M V$ with $V$ being a $(d-1)(d+2)/2$ dimensional vector with the independent components of $T_{\mu\nu}$ as its components. If we diagonalize the matrix $M$ with a matrix, say $U$, then it is possible to write 
\be\label{diagonalconstT}
\Delta^{(2)}S=(UV)^TM_d(UV)=\sum_{i}\lambda_i(UV)_i\,.
\ee 
We must now demand that the eigenvalues $\lambda_i$ are negative so that  $\Delta^{(2)}S< 0$ and hence the relative entropy is positive. This leads to the following inequalities for $n_1,n_2$ \cite{constrgrav}:
\begin{eqnarray}
n_1+2(d-1)n_2 &\geq& 0 \,,\\
2d+1-4(d+1)n_1-4(d^2-1)n_2 &\geq& 0\,,\\
d+2-4(d+1)n_1-4d(d^2-1)n_2 &\geq& 0\,,
\end{eqnarray}
 which lead to the enclosed region shown in figure 1.

\begin{figure}[ht]
\centering
\boxed{\includegraphics[scale=1.2]{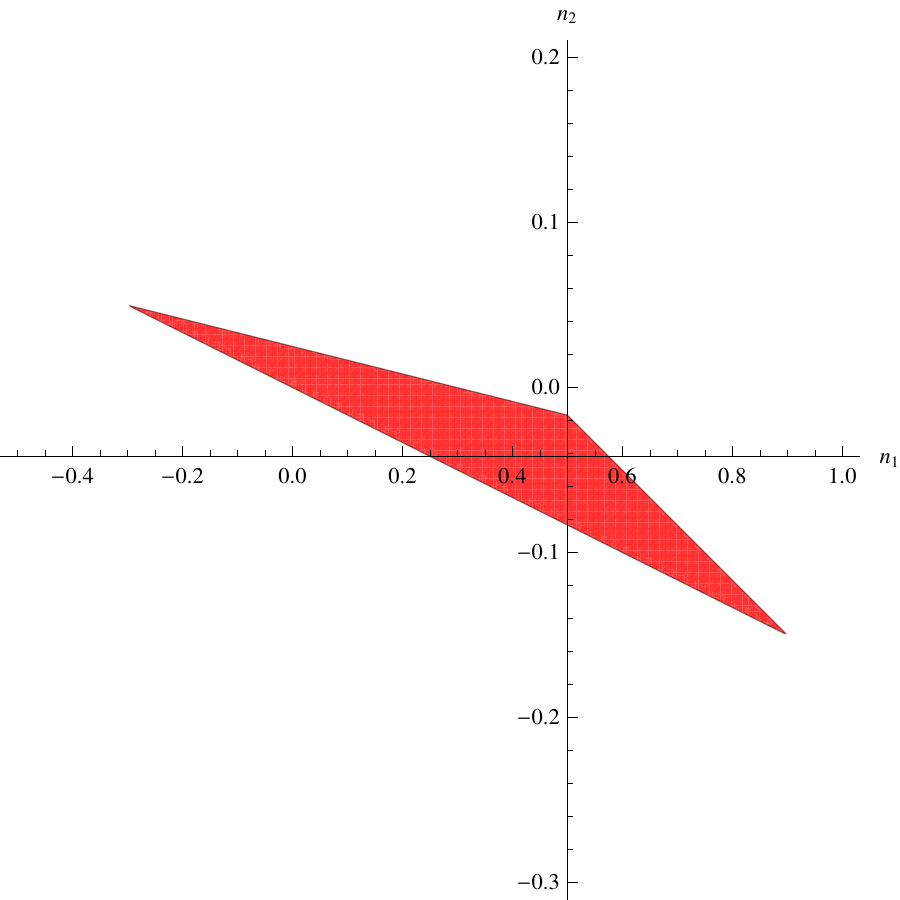}}
\caption{(colour online) For $d>2$ we get the allowed $n_1,n_2$ region for constant stress tensor to be the triangle above \cite{constrgrav}. The Einstein value $(n_1,n_2)=(\frac{1}{2},-\frac{1}{8(d-1)})$ is at the origin of the plot. }
\end{figure}

The fact that we get a region of finite area in the $n_1,n_2$ parameter space which encloses the Einstein point is encouraging. However, this was for a constant stress tensor. What happens to this constrained region for a non-constant stress tensor? In order to answer this question we will need to find the Green's function for eq.(\ref{gf}). For concreteness we will focus on $d=4$ henceforth. One can hope that the allowed parameter space shrinks on considerations of non-constant stress tensor (perhaps down to the Einstein point). Furthermore, we will be introducing more parameters to account for the derivatives of the stress tensor and we can ask if these additional parameters are themselves constrained. 

This paper is organized as follows. In sections 2 we will explain in some detail how the Green's function to find $z_1$ is derived. Using this we will consider the change in the relative entropy due to non-constant stress tensors in section 3. In section 4 we will show that for Einstein values of the nonlinear parameters relative entropy is positive, and also derive constraints on these parameters using the positivity. In section 5, we will generalize the analysis in arbitrary dimensions. We will conclude in section 6. There are three appendices with useful details of intermediate steps.

\section{A systematic way to find $z_1$}
Let us start with a four dimensional theory living on $R^{3,1}$. The dual gravity theory lives on $AdS_5$ whose metric in Poincare coordinates is given by, 
\begin{equation}
ds^2 = \frac{L^2}{z^2}(dz^2 + d{\vec{x}}^2 - dt^2)\,.
\end{equation}
We want to compute the entanglement entropy of a ball of radius unity\footnote{The discussions in this section can be easily generalized for an arbitrary radius $R$. The rest of the paper assumes a generic radius. } in the 3+1-dimensional boundary. Without any loss of generality we can assume that the ball is centered at the origin and so the entangling surface is given by, $\vec x^2 = r^2= 1$. The corresponding minimal surface in $AdS_5$ is given by the equation, $z^2 + \vec x^2 = 1$. If we change the bulk geometry then this minimal surface will change. If $\epsilon$ denotes the strength of the perturbation then we can write, 
\begin{equation}
g_{\mu\nu} = g^{(0)}_{\mu\nu} + \epsilon g^{(1)}_{\mu\nu} + \epsilon^2 g^{(2)}_{\mu\nu} + \cdots
\end{equation}
and 
\begin{equation}
z (\vec x) = z_0 (\vec x) + \epsilon z_1 (\vec x) + \epsilon^2 z_2 (\vec x) + \cdots\,,
\end{equation}
where $z(\vec x)$ is the equation of the new minimal surface and $z_0$ is the unperturbed minimal surface given by the equation $z^2 + \vec x^2 = 1$. Our aim is to find the first order perturbation given by $z_1$. In order to do that we have to find out the change in the area functional due to the change in the metric up to second order and impose the minimality constraint. The process is straightforward and leads to the following differential equation for $z_1(\vec x)$, 
\be\label{diffeq}
z_0\left(\partial ^2\left(z_0z_1\right)-x^ix^j\partial _i\partial _j\left(z_0z_1\right)\right)=
J\,.
\ee
Here, $J$ is a source function which depends on the details of the metric deformation. The explicit form of $J$ is shown in (\ref{gf}) for deformations with a constant boundary stress tensor, and in Appendix \ref{solnonconstT} we compute it for the non-constant case. Now the key point is to note that the induced metric on the zeroth order minimal surface $z_0$ is that of an Euclidean $AdS_3$, with the entangling surface $r=1$ as its conformal boundary. This will be clear from the coordinate transformations described in the following subsections. For the time being let us write the differential equation in the form,
\begin{equation}\label{me1}
(- \Delta_{H^3} + 3) z_1 =  J\,,
\end{equation}
where $\Delta_{H^3}$ is the scalar Laplacian on $AdS_3$. The origin of this $AdS_3$-Laplacian can be understood in the following way. First of all we are dealing exclusively with time-independent deformations of the bulk metric and so we can again parametrize the minimal surface by an equation of the form $z= z(\vec x)$. The first order fluctuation $z_1$ can then be thought of as a scalar field propagating on the unperturbed minimal surface. Since, the unperturbed minimal surface $z_0$ is an Euclidean $AdS_3$ we automatically get the Laplacian on $AdS_3$. 

Eq. (\ref{me1}) is the equation of a scalar field propagating on $AdS_3$ with $m^2 = 3$, where $m$ is the mass of the scalar field. 
The general solution can be written as, 
\begin{equation}
z_1 = \int G_{bulk-bulk} J + f_{hom}\,,
\end{equation}
where $G_{bulk-bulk}$ is the bulk to bulk propagator for a massive scalar field on $AdS_3$ and $f_{hom}$ is the solution of the homogeneous equation subject to the proper boundary condition. A scalar field with $m^2 = 3$ corresponds to an irrelevant operator in the $CFT_2$ and so the homogeneous solution grows towards the boundary of the $AdS_3$ which is the entangling surface in our case. We need to set this mode to zero because $z_1$ vanishes on the entangling surface.

\subsection{The Green's function for $z_1$}
The general solution for $z_1$ can be written as,
\begin{equation}\label{z1sol}
z_1(x) = \int d\mu_{\hat{x}} G(x,\hat{x}) J(\hat{x})\,,
\end{equation}
where $d\mu_{\hat x}$ is the Riemannian volume element on $AdS_3$ and we denote the intrinsic coordinates collectively by $x$. In terms of embedding coordinates, an $AdS_3$ of unit radius is described by a hyperboloid in Minkowski space $R^{3,1}$ given by,
\begin{equation}
X_1^2 + X_2^2 + X_3^2 - X_4^2 = -1\,,
\end{equation}
where $\vec X$ are the coordinates\footnote{This $R^{3,1}$ should not be confused with the boundary of the Poincare patch where the CFT lives.} of $R^{3,1}$. One can introduce  intrinsic coordinates on $AdS_3$ by, 
\begin{equation}
X_1 = \sin\theta  \ \cos\phi  \ \sinh\eta , \
X_2 = \sin\theta \ \sin\phi \ \sinh\eta, \
X_3 = \cos\theta \ \sinh\eta, \
X_4 = \cosh\eta\,.
\end{equation}
In terms of the boundary(CFT) spherical polar coordinates, $\eta=\tanh^{-1}(r)$. The angular coordinates $\theta$ and $\phi$ are the usual ones. The metric then takes the following simple form in terms of the intrinsic coordinates,
\begin{equation}
ds^2 = d\eta^2 + \sinh^2\eta (d\theta^2 + \sin^2\theta d\phi^2)\,.
\end{equation}
It is easy to check that this is the induced metric on the minimal surface $z^2 + \vec x^2 =1$, if we write,
\begin{equation}
z = \sqrt{(1 - \vec x^2)} = \sqrt {(1 - r^2)}=  \ \text{sech}\eta, \  d\vec x^2 = dr^2 + r^2 (d\theta^2 + \sin^2\theta d\phi^2), \  dt=0\,.
\end{equation}
and  substitute it back into the metric of $AdS_5$ written in Poincare coordinates.

Given two points $x$ and $\hat x$ on $AdS_3$ the geodesic distance $d(x,\hat x)$ is given by the relation,
\begin{equation}
\cosh \ d(x,\hat{x}) = - X\cdot\hat{X}\,,
\end{equation}
where $X$ and $\hat X$ are the respective embedding coordinates, corresponding to $x$ and $\hat{x}$. In (\ref{z1sol}), $G(x,\hat x)$ is the Green's function i.e., as mentioned before, the bulk to bulk  propagator for a scalar field with $m^2=3$. It is given by \cite{dhoker},
\be
G(x,\hat{x}) = \frac{1}{4\pi} \frac{e^{-2d(x,\hat{x})}}{\sinh \ d(x,\hat{x})}\,.
\ee

Let us now work out a simple example. Consider the situation where a constant stress tensor has been switched on in the field theory. This stress tensor gives rise to metric fluctuation in the bulk and we want to compute the new minimal surface resulting from that. For a constant stress tensor, the source $J$ is given by,
\begin{equation}
J = - {z_0}^5 (T + 3T_x)\,,
\end{equation}
where $T_x = x^i x^j T_{ij}$. $x^i$, where $i$ runs from $1$ to $3$, denotes the cartesian coordinates of the boundary $R^3$.
Let us now see how things simplify if we take an isotropic stress tensor. The isotropic stress tensor is given by,
\begin{equation}
T_{ij} = \frac{1}{3} T \delta_{ij} \,,
\end{equation}
where $T$ is a constant. $T_x = x^i x^j T_{ij} = \frac{T}{3} r^2 = \frac{T}{3} \tanh^2\eta$ .  Let us take the point $x$ in (\ref{z1sol}) to be $\eta=0$. In terms of embedding coordinates this is the point $X_1 = X_2 = X_3 = 0, X_4 = 1$. So this is the lowest point of the hyperboloid. At this point our calculation gets enormously simplified, since, we get $\cosh d=\cosh \hat{\eta}$, where $\hat{\eta}$ is the intrinsic coordinate corresponding to $\hat{x}$. This gives the solution,
\begin{equation}
z_1(0) = \int d\hat{\Omega}_2 \int_{0}^{\infty} d\hat{\eta} \sinh^2\hat{\eta} \frac{1}{4\pi} \frac{e^{-2\hat{\eta}}}{\sinh \ \hat{\eta}}J(\hat{\eta}) \hspace{1cm}\text{where}\hspace{0.5cm}J(\hat{\eta})=\text{sech}^5\hat{\eta}(1+\tanh^2 \hat{\eta})\,.
\end{equation}
We have set the homogeneous solution to zero because that grows at the boundary and so is inconsistent with the boundary condition $z_1=0$.
Then the above integral gives,
\be
z_1(0)=-\frac{T}{10}\,.
\ee
This is indeed what we would get in $d=4$ from the solution given in \cite{myers} for an isotropic constant $T_{\mu\nu}$ with $R=1$ and at $r=0$. So the Green's function gives the correct solution in this case.

Note that for this particular example, the source is rotationally invariant. Since the Green's function and measure are both rotationally invariant, $z_1$ is also rotationally invariant. That means $z_1$ is a function of $\eta$ only. We can use the fact that $AdS_3$ is a homogeneous space of the Lorentz group $SO(3,1)$ and so given two points on $AdS_3$, there exists an $SO(3,1)$ group element which maps one point to the other. This allows us to simplify the calculation for a general $\eta$. In the following subsection, we will show how to find the solution for an arbitrary point, even for an arbitrary source which is not rotationally invariant.

\subsection{General solution for an arbitrary source}\label{gensol}
In this section we shall do the integrals in a way which makes life easier even for arbitrary source and it can be easily automated.  Suppose we want to compute $z_1$ at an arbitrary point, $P$, which is not necessarily the origin $\eta=0$. $AdS_3$ is the homogeneous space of the group of isometries $SO(3,1)$. 
Our strategy will be the following. We shall make a coordinate transformation such that in the new coordinate system the point $P$ is at the origin, $\eta'=0$, where the primed coordinates are the new transformed coordinates. This coordinate transformation can be chosen to be an isometry and so the form of the differential operator does not change. The Green's function has exactly the same form in the new coordinates. What makes life easier is the fact that the geodesic distance which appears in the Green's function integral is now given by $\hat{\eta}'$ (where $\hat{\eta}$ is the integration variable). So we do not have to deal with complicated expressions for the geodesic distance in the integrand. This makes things much easier.  

We have the Cartesian coordinates $(x_1,x_2,x_3)$ and the polar coordinates $(r,\theta,\phi)$ related by,
\begin{equation}
x_1 = r \ \sin\theta \ \cos\phi, \ x_2 = r \ \sin\theta \ \sin\phi, \ x_3= r \ \cos\theta\,.
\end{equation} 
We can go to the standard polar coordinates $(\eta,\theta,\phi)$ on $H_3$ by the transformation,
\begin{equation}
r= \tanh\eta\,.
\end{equation}
The embedding coordinates of $H_3$ are given by,
\begin{equation}
X_1 = \sin\theta  \ \cos\phi  \ \sinh\eta , \
X_2 = \sin\theta \ \sin\phi \ \sinh\eta, \
X_3 = \cos\theta \ \sinh\eta, \
X_4 = \cosh\eta\,.
\end{equation}
So comparing these equations we get,
\begin{equation}
x_i = \frac{X_i}{X_4}, \ i=1,2,3\,.
\end{equation}
The coordinate transformation can be written in a simple form in terms of embedding coordinates. The coordinate transformations can be built out of three matrices, two of which are rotations and one boost. This is the most general coordinate transformation required. The matrices are given by, \\
\begin{align}
&K_{34}(\alpha)=
\begin{pmatrix}
1&0&0&0 \\
0&1&0&0 \\
0&0&\cosh\alpha & \sinh\alpha \\
0&0&\sinh\alpha & \cosh\alpha
\end{pmatrix}\,,  \\ 
&R_{13}(\beta) =
\begin{pmatrix}
\cos\beta & 0&\sin\beta&0 \\
0&1&0&0 \\
-\sin\beta & 0&\cos\beta&0 \\
0&0&0&1
\end{pmatrix}\,, \
R_{12}(\gamma) = 
\begin{pmatrix}
\cos\gamma & -\sin\gamma &0&0 \\
\sin\gamma & \cos\gamma &0&0 \\
0&0&1&0 \\
0&0&0&1
\end{pmatrix}\,.
\end{align}
In terms of the embedding coordinates the coordinate transformation can be written as,
\begin{equation}
\begin{pmatrix}
X'_1 \\
X'_2 \\
X'_3 \\
X'_4
\end{pmatrix}
= K^{-1}_{34} R^{-1}_{13} R^{-1}_{12} 
\begin{pmatrix}
X_1 \\
X_2 \\
X_3 \\
X_4
\end{pmatrix}\,.
\end{equation}
In the old coordinates, the origin $\eta=0$ corresponds to the point $(0,0,0,1)$ in the embedding coordinates, $(X_1,X_2,X_3,X_4)$. It is easy to see that the point $(X'_1=0,X'_2=0,X'_3=0,X'_4=1)$ corresponds to the point $(X_1= \sin\beta \ \cos\gamma \ \sinh\alpha, X_2 = \sin\beta \ \sin\gamma \ \sinh\alpha, X_3= \cos\beta \sinh\alpha, X_4 = \cosh\alpha)$ in the old embedding coordinates. So we can place the origin of our new coordinates at any point on $H_3$ by varying $(\alpha,\beta,\gamma)$. So we have achieved our goal. Since the coordinate transformation is an element of $SO(3,1)$ we have,
\begin{equation}
{X'_1}^2 + {X'_2}^2 + {X'_3}^2 - {X'_4}^2 = -1\,.
\end{equation}
So we can introduce new intrinsic coordinates given by,
\begin{equation}\label{newintrs}
X'_1 = \sin\theta'  \ \cos\phi'  \ \sinh\eta' , \
X'_2 = \sin\theta' \ \sin\phi' \ \sinh\eta', \
X'_3 = \cos\theta' \ \sinh\eta', \
X'_4 = \cosh\eta'\,.
\end{equation}
In terms of the new intrinsic coordinates the point, $(\eta=\alpha, \theta=\beta, \phi=\gamma)$ goes to, $\eta'=0$. Since $z_1$ is a scalar its value at the point $(\eta=\alpha, \theta=\beta, \phi=\gamma)$, is the same as its value at the point, $\eta'=0$. The new intrinsic coordinates are some complicated functions of the old ones, but since its an isometry, the form of the differential operator does not change and so is the Green's function. We want to compute $z_1$ at the origin of this new coordinates. So the answer is given by\footnote{The integration variables in (\ref{gensolarbitJ}) should have been $(\hat{\eta}',\hat{\theta}',\hat{\phi}')$. The primes have been removed for notational simplification. },
\begin{equation}\label{gensolarbitJ}
z_1(0') = z_1(\eta,\theta,\phi) = \int\ d\hat{\Omega}_2 \int_{0}^{\infty} d\hat{\eta} \sinh^2\hat{\eta} (\frac{1}{4\pi}\frac{e^{-2\hat{\eta}}}{\sinh\hat{\eta}}) J'(\hat{\eta},\hat{\theta},\hat{\phi})\,.
\end{equation}
We have used the fact that the geodesic distance from the origin is given by $\hat{\eta}$. We can see that all the dependence on $(\eta,\theta,\phi)$ arises through the source. Let us now say a few words about the source. Since the source is a scalar the functional form of $J'$ can obtained via, 
\begin{equation}
J'(\eta',\theta',\phi') = J(\eta(\eta',\theta',\phi'), \theta(\eta',\theta',\phi'),\phi(\eta',\theta',\phi'))\,.
\end{equation}
The explicit coordinate transformation between the two sets of intrinsic coordinates is complicated. The simpler thing to do is to first express the source $J(\eta,\theta,\phi)$ in terms of Cartesian coordinates $x_i = \frac{X_i}{X_4}$. We can now use the known functional dependence of old embedding coordinates in terms of the new to express the source in new intrinsic coordinates (by using (\ref{newintrs})). 

One can show that this method gives the correct answer for an arbitrary constant stress tensor\footnote{See appendix \ref{const T} for a detailed calculation in the case of arbitrary constant source.}. There are many ways to do it, but we have shown the one that we will use for the non-constant stress tensor. We now move on to the non-constant stress tensor, in which we use the above treatment to find the relative entropy correction.

\section{Non-constant stress tensor}
\subsection{Deformation of the metric}
For a space-dependent perturbation, let us restrict our attention to stress tensors whose variations are small compared to the the size of the entangling region. To be precise, we want \footnote{$\mathcal{O}(\cdots)$ refers to all possible contractions of the tensors appearing in the argument of $\mathcal{O}$}
\be\label{contraint1}
\mathcal{O}\left(R^2 \partial T \partial T )\ll \mathcal{O}( R^2 \partial\partial T\right)
\ee
and
\be\label{constraint2}
\mathcal{O}\left(R^2 \partial\partial T \right)\ll \mathcal{O}\left( R\partial T\right)\,,
\ee 
and higher derivatives are similarly suppressed. So we will not consider beyond two derivatives of $T_{\mu\nu}$ in our calculations. Note that, $T_{\mu\nu}$ must satisfy the traceless ($T_{\mu}^\mu =0$) and divergenceless ($\partial_\mu T^\mu_\nu = 0$) conditions.
The boundary metric up to two derivatives in the quadratic correction, looks like
\be\label{metric}
\frac{z^2}{L^2} \ g_{\mu  \nu }=\eta _{\mu  \nu }+z^4\left(T_{\mu  \nu }-\frac{1}{12}z^2\square T_{\mu  \nu }\right)+z^8\left(n_1T_{\mu  \alpha }T_{\nu }{}^{\alpha }+n_2\eta _{\mu  \nu }T_{\alpha  \beta }T^{\alpha  \beta }+z^2\mathcal{T}_{\mu  \nu }\right)\,.
\ee  
The last term is given by,
\begin{align}\label{n3n4}
\mathcal{T}_{\mu  \nu }=&n_3\left(T_{\mu  \alpha }\square T_{\nu }{}^{\alpha } + T_{\nu  \alpha }\square T_{\mu }{}^{\alpha }\right)+n_4\eta _{\mu  \nu }T_{\alpha  \beta }\square T^{\alpha  \beta }+n_5\partial _{\mu }T_{\alpha  \beta }\partial _{\nu }T^{\alpha  \beta }+n_6\partial _{\alpha }T_{\mu  \beta }\partial ^{\beta }T_{\nu }{}^{\alpha }+n_7\partial _{\mu }\partial _{\nu }T_{\alpha  \beta }T^{\alpha  \beta }\nonumber\\&+n_8\partial _{\alpha }T_{\mu  \beta }\partial ^{\alpha }T_{\nu }{}^{\beta }+n_9\left(\partial _{\mu }T_{\alpha  \beta }\partial ^{\beta }T_{\nu }{}^{\alpha }+\partial _{\nu }T_{\alpha  \beta }\partial ^{\beta }T_{\mu }{}^{\alpha }\right)+n_{10}\eta _{\mu  \nu }\partial _{\alpha }T_{\beta  \gamma }\partial ^{\alpha }T^{\beta  \gamma }+n_{11}\partial _{\alpha }T_{\gamma  \beta }\partial ^{\beta }T^{\gamma  \alpha }\eta_{\mu\nu}\nonumber\\&+n_{12}\left(T^{\beta  \alpha }\partial _{\alpha }\partial _{\mu }T_{\nu  \beta }+T^{\beta  \alpha }\partial _{\alpha }\partial _{\nu }T_{\mu  \beta }\right)+n_{13}T^{\alpha  \beta}\partial _{\alpha }\partial _{\beta }T_{\mu  \nu }  \,.
\end{align} 
In other words, the leading correction comes from $T_{\mu\nu}$ and subleading from $\square T_{\mu\nu}$.  It was shown in \cite{myers} that the entropy change corresponding to both of these, at linear order, satisfies,
\be
\Delta S=\Delta H\,.
\ee
So we must look at the corrections coming from the $\mathcal{O}(TT)$, $\mathcal{O}(\partial T\partial T)$ and $\mathcal{O}(T\partial \partial T)$ terms. We expect these contributions to be negative. From eqs.(\ref{contraint1}) and (\ref{constraint2}) it is clear that there will be no correction from any higher derivative terms. So the above form of metric will suffice. 

The parameters $n_1, \cdots, n_{13}$ appearing in the metric have to be fixed from the Einstein equations. From our knowledge of the constant perturbation, we already know that $n_1=1/2$ and $n_2=-1/24$. Using a convenient functional form of $T_{\mu\nu}$ satisfying the traceless and divergence-less conditions, we compute $R_{\mu\nu}-\frac{1}{2}R g_{\mu\nu}-6g_{\mu\nu}$ in terms of $n_{3},\cdots,n_{13}$ and set all the components to zero. Then we see that for any such choice of stress tensor, the Einstein equations  will be satisfied only if the parameters in (\ref{n3n4})  take the values, 
\begin{align} \label{n3n4values}
& n_3=-\frac{1}{24},n_4=\frac{1}{180},n_5=-\frac{1}{180},n_6=-\frac{1}{60},n_7=\frac{1}{360},\nonumber\\&
n_8=0,n_9=\frac{1}{120},n_{10}=\frac{1}{720},n_{11}=0,n_{12}=-\frac{1}{120},n_{13}=\frac{1}{60}\,.
\end{align}
We will begin by keeping all the above parameters to be arbitrary. This will make the correction $\Delta^{(2)}S$ dependent on these 13 parameters. So, we have a 13-dimensional parameter space, instead of 2. However, it is always possible to constrain a certain subspace of some (say two) parameters, by fixing the rest at the Einstein values. Of course, we are primarily interested in the subspace of $n_1$ and $n_2$, which correspond to the zeroth order in the quadratic derivative expansion.

\subsection{The correction to $z_1$ and area functional}
To get the change of minimal surface, $z_1$, we need to find the source function. The source term for non-constant stress tensor is obviously different from the constant case. We can find it by calculating the area functional and minimizing it w.r.t. $z_1$. We have shown the steps in appendix \ref{solnonconstT}. The source term works out to be\footnote{The nonlinear terms in eq.(\ref{n3n4}) will not modify the source term since they are second order in the perturbation while the source is first order.},
\be\label{source}
J=-\frac{z_0^5}{2}\left(2T+6T_x-\frac{z_0^2}{3}\left(\partial^2T+2\frac{x^ix^j}{R^2}\partial^2T_{ij}\right)+x^i\partial_i T+2x^i\partial_0{T}_{0j}+ \frac{1}{R^2}x^ix^jx^k\partial_k T_{ij}\right)\,,
\ee
where we have retained terms only upto two derivatives of the stress tensor. Note that we are working in a time-independent background. So, time derivatives of the stress tensor do not appear.

Now we use the formula (\ref{gensolarbitJ}) with the above source to find $z_1$. The trick is to go to Fourier space, so that the source is nothing but an exponential. Then the actual source (\ref{source}) is written in term of derivatives of the exponential to find the actual solution. The method is shown in detail in appendix \ref{const T}  and appendix \ref{solnonconstT}, for constant $T_{ij}$ and non-constant $T_{ij}$ respectively. Here we just quote the result,
\begin{align}\label{z1}
z_1&=-z_0{}^3R^2\left(\frac{T+T_x}{10}+\frac{1}{12}\left(x^i\partial _iT + x^ix^jx^k\frac{\partial _kT_{i j}}{R^2}\right)\right.\nonumber\\&\left.+\frac{1}{28} \left(x^ix^j\partial _i\partial _jT + x^ix^jx^kx^l\frac{\partial _i\partial _jT_{k l}}{R^2}\right)-\frac{k^2 \left(R^2-r^2\right)}{168}\left(\partial ^2T+x^ix^j\frac{\partial ^2T_{i j}}{R^2}\right)\right)\,.
\end{align}
Here all $T_{ij}$-s and their derivatives are evaluated at the origin and this will be the case from here onwards, unless and otherwise mentioned. The next step is to evaluate the area functional in terms of $T_{ij}$-s and derivatives at the origin.

We already know how the area depends on only the $T^2$ terms (\ref{del2SconstT}). Now the next order should contain one-derivative of $T_{ij}$. This means we need to integrate over terms like $T_{ij}x^k \partial_k T^{ij}$, $T_{ij}x^ix^jx^k\partial_kT$ and so on. All these terms have an odd number of $x$-s. If we assume that $T_{ij}$ is a smooth function, this integral must vanish,
 
\be
\int d^{d-1}x T_{ij} \ (x^mx^n...\text{odd no.}) \ \partial_kT^{pq}= T_{ij}\partial_kT^{pq}\int d^{d-1}x x^m x^n... =0\,.
\ee

Thus, we have to evaluate the area upto two derivatives of $T_{ij}$. This was the reason why the metric (\ref{metric}) was constructed upto two derivatives of $T_{\mu\nu}$.  Since  the $z_1$ solution depends on derivatives of $T_{ij}$ evaluated at origin, we must do the same for the area functional. This means all $T_{ij}(\vec{x})$ appearing in the area formula must be Taylor expanded around the origin first, and then integrated. The details of this straightforward but tedious calculation has been shown in the appendix \ref{areafunc}. Here we quote the result,
\begin{align}\label{A2}
&\int d^3x \sqrt{h}\!=\frac{4\pi  L^3R^{10}}{31185}\left[\left(10-12 n_1+2160 n_{11}+720 n_6+1440 n_9\right)\left(\partial _iT_{j k}\partial ^kT^{j i}\right)+48 \left(7 n_2+45 n_4\right.\right.\nonumber\\&
 \left.+15 n_7\right)T \partial ^2T+\left(-120 n_1-672 n_2-1440 n_3-4320 n_4-1440 n_7\right)T^{0 i}\partial ^2T_{0 i}+\left(-12+720 n_{13}\right) T^{i j}\partial _i\partial _jT\nonumber\\&
+\left(-55+120 n_1+2160 n_{10}+336 n_2+720 n_5+720 n_8\right) \left(\partial _iT_{j k}\right){}^2+\left(12 n_1-2160 n_{11}-1440 n_9\right) \partial _iT_{0 j}\partial ^jT^{0 i}\nonumber\\&
+\left(5+2160 n_{10}+336 n_2+720 n_5\right) \left(\partial _iT\right){}^2+\left(120 n_1+336 n_2+1440 n_3+2160 n_4+720 n_7\right) T^{i j}\partial ^2T_{i j}\nonumber\\&
\left.+\left(-120 n_1-4320 n_{10}-672 n_2-1440 n_5-720 n_8\right) \left(\partial _iT_{0 j}\right){}^2\right]\ +\ \mathcal{O}(T T)\,.
\end{align}
In $d=4$, the metric (\ref{metric}) satisfies Einstein's equations for $n_1=1/2$ and $n_2=-1/24$ and the values given in (\ref{n3n4values}). For these values the expression above simplifies to,
\be\label{A2enst}
\int d^3x \sqrt{h}=-\frac{8\pi L^3 \left(5 \left(\partial _iT\right){}^2+15 \left(\partial _iT_{0 j}\right){}^2+3 \partial _iT_{0 j}\partial ^jT_0{}^i+5 \left(\partial _iT_{j k}\right){}^2-2 \partial _iT_{k j}\partial ^kT^{i j}\right) R^{10}}{31185}+ \mathcal{O}(T T)\,.
\ee
Very interestingly, all the `cross' tensor structures, i.e. $T^{ij}\partial^2T_{ij}$, $T^{ij}\partial_i\partial_j T$, $T \partial^2T$ and $T_0^i\partial^2T_{0i}$ vanish at these values. But first, let us see what we can imply from the result (\ref{A2}).
Let us  add $(\ref{A2})$ to the previous result $(\ref{del2SconstT})$. 
This gives the total subleading correction to the entropy in terms of $T_{\mu\nu}$ and its derivatives at the origin.
\begin{align}\label{d2Sgen}
&\Delta^{(2)}S=\frac{8\pi^2 L^3 R^8}{4725\ell_P^3}\left(-160(n_1+6 n_2) \left(T_{i 0}\right){}^2+8 (-9+20 n_1+60 n_2) \left(T_{i j}\right){}^2+8 (1+60 n_2)T^2\right)+\nonumber\\&
+\frac{8\pi^2   L^3R^{10}}{31185\ell_P^3}\left[\left(10-12 n_1+2160 n_{11}+720 n_6+1440 n_9\right)\left(\partial _iT_{j k}\partial ^kT^{j i}\right)+48 \left(7 n_2+45 n_4+15 n_7\right) T \partial ^2T\right.\nonumber\\&
+\left(-120 n_1-672 n_2-1440 n_3-4320 n_4-1440 n_7\right)T^{0 i}\partial ^2T_{0 i}+\left(-12+720 n_{13}\right) T^{i j}\partial _i\partial _jT\nonumber\\&
+\left(-55+120 n_1+2160 n_{10}+336 n_2+720 n_5+720 n_8\right) \left(\partial _iT_{j k}\right){}^2+\left(12 n_1-2160 n_{11}-1440 n_9\right) \partial _iT_{0 j}\partial ^jT^{0 i}\nonumber\\&
+\left(5+2160 n_{10}+336 n_2+720 n_5\right) \left(\partial _iT\right){}^2+\left(120 n_1+336 n_2+1440 n_3+2160 n_4+720 n_7\right) T^{i j}\partial ^2T_{i j}\nonumber\\&
\left.+\left(-120 n_1-4320 n_{10}-672 n_2-1440 n_5-720 n_8\right) \left(\partial _iT_{0 j}\right){}^2\right]\,. 
\end{align}

\section{Constraints on the nonlinear parameters}
For a unitary theory, we expect the above quantity (\ref{d2Sgen}) to be negative. In other words, only those values of $n_1, \cdots, n_{13}$, for which the above quantity is manifestly negative, are viable for unitarity. But first we need to make sure that this is indeed the case for Einstein values of the thirteen parameters. For these values, i.e. $n_1=1/2$, $n_2=-1/24$ and (\ref{n3n4values}), the total subleading correction simplifies to,
\begin{align}\label{d2Enst}
\Delta^{(2)}S=&-16\pi^2  R^{10}\frac{L^3}{\ell_P^3}\left[\frac{ 6 T^2+20 \left(T_{i 0}\right){}^2+6 \left(T_{i j}\right){}^2}{4725R^2}\!\right.\nonumber\\&\left.+\frac{  \left(5 \left(\partial _iT\right){}^2+15\! \left(\partial _iT_{0 j}\right){}^2+3 \partial _iT_{0 j}\partial ^jT_0{}^i+5\! \left(\partial _iT_{j k}\right){}^2-2 \partial _iT_{k j}\partial ^kT^{i j}\right) }{31185}\right]\,. 
\end{align}
We already knew that the $\mathcal{O}(TT)$ part of this is negative. Fortunately, the $T\partial\partial T$ `cross' structures, which are not manifesltly positive definite,  do not appear at these values. However there are still two terms, namely $\partial_iT_{0j}\partial^jT^{0i}$ and $\partial_iT_{jk}\partial^kT^{ij}$, which are not positive definite. So to ensure the negativity of the expression (\ref{d2Enst}) we have to express it as a sum of squares. As we did in (\ref{diagonalconstT}), we can write it as a matrix inner product, $V^TMV$, where $V$ is a column vector. The elements of $V$ are of the form, $\partial_iT_{j\mu}$ and they are linearly independent of each other, made sure by the constraints,
\be
\partial^iT_{ij}=0\,,\hspace{1cm}\partial^iT_{i0}=0\,\hspace{1cm}\text{and}\hspace{1cm}\partial_iT_{jk}=\partial_iT_{kj}\,.
\ee
We choose $V$ to be,
\begin{multline}
V=\left\{\partial _1T_{02},\partial _1T_{03},\partial _2T_{01},\partial _2T_{02},\partial _2T_{03},\partial _3T_{01},\partial _3T_{02},\partial _3T_{03},\partial _1T_{12},\partial _1T_{13},\partial _1T_{22},\right.\\
\left.\partial _1T_{32},\partial _1T_{33},\partial _2T_{11},\partial _2T_{21},\partial _2T_{31},\partial _2T_{32},\partial _2T_{33},\partial _3T_{11},\partial _3T_{21},\partial _3T_{31},\partial _3T_{22},\partial _3T_{32}\right\}\,.
\end{multline}
Now we diagonalize the $23\times 23$ matrix $M$ with a matrix $U$. Then we can write,
\be\label{diagsumEnst}
\Delta^{(2)}S=V^TMV=V^TU^TM_dUV=(UV)^TM_d(UV)=\sum_{i=1}^{23}\lambda_i {(UV)_i}^2\,.
\ee
Here $\lambda_i$ are the eigenvalues of the matrix $M$. For the Einstein values of $n_1$ and $n_2$ they are\footnote{The superscripts on the eigenvalues indicate degeneracy} (in units of $8\pi^2  R^{10}L^3/\ell_P^3$),
\begin{align}
\lambda=&\{-0.000769^{\ \times 3},-0.00115^{\ \times 4},-0.00346,-0.000769^{\ \times 2},-0.000384,-0.000577^{\ \times 3},-0.000256^{\ \times 3},\nonumber\\
&-0.00207^{\ \times 3},-0.000428^{\ \times 3}\}\,.
\end{align}
Since all are negative, the sum in (\ref{diagsumEnst}) is manifestly a negative definite quantity. Pleasingly, this confirms our expectation that, at this nonlinear order, the CFT dual to Einstein gravity is unitary.

The main question of interest is if the 13-dimensional parameter space is bounded by the inequality $\Delta^{(2)}S\le 0$. To begin with let us ask what is the influence of the non-constant stress tensor on the ($n_1, n_2$) parameter space. Let us begin by setting all the rest of the $n_i$'s to their Einstein  values (\ref{n3n4values}). A problem that will arise in this case is that the $\mathcal{O}(T\partial\partial T)$ `cross' terms will have non-zero coefficients. The above method of diagonlizing the coefficient matrix will fail because the cross terms cannot give a sum of squares. Even if we try to combine them with the $\mathcal{O}(TT)$ terms to complete the square (i.e. $\mathcal{O}(TT)+\mathcal{O}(T\partial\partial T) \rightarrow \mathcal{O}((T+\partial T)(T+\partial T))$) then some new terms $\mathcal{O}(\partial\partial T\partial\partial T)$ will appear. Since we don't know how such terms actually appear in $\Delta^{(2)}S$, it is not a good idea to do it this way.

\begin{figure}[ht]
\centering
\boxed{\includegraphics[scale=1.1]{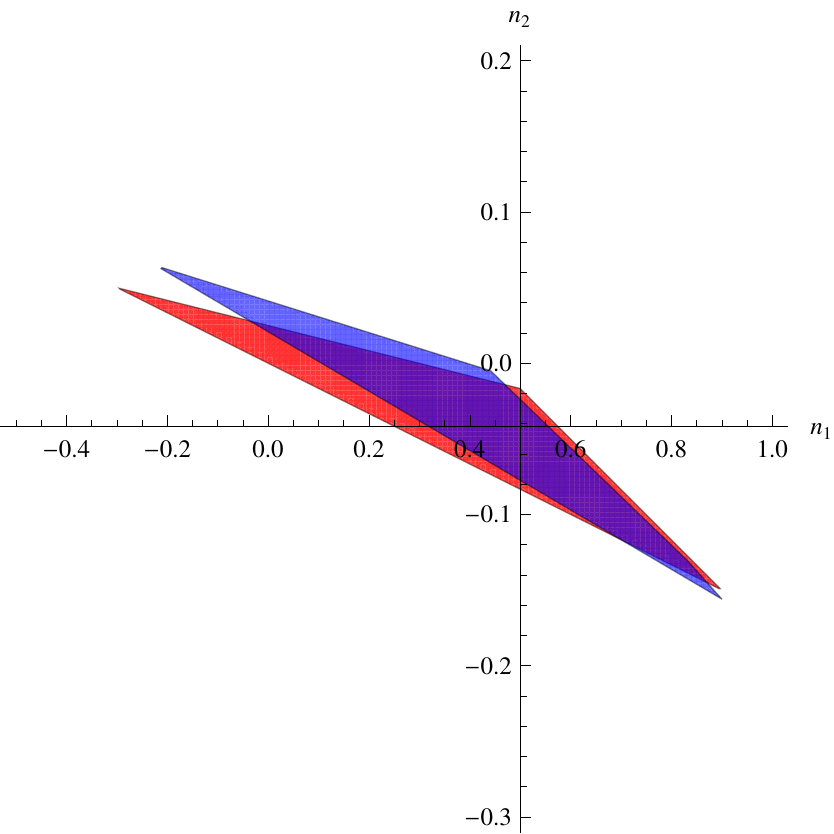}} 
\caption{(colour online)  The blue (upper) triangle is the new region we have obtained from the $\mathcal{O}(\partial T\partial T)$ terms. This constraint is independent of the old one (shown in red) obtained from  $\mathcal{O}(T T)$ terms. The intersecting part is the net allowed region for $n_1$ and $n_2$. }\label{oldnew}
\end{figure}

The best way forward is to make the stress tensor vanish at the origin, i.e.,
\be
T_{\mu\nu}(\vec{x}=0)=0\,.  
\ee
Note that this does not violate our assumptions (\ref{contraint1}) and (\ref{constraint2}). Since all we want is to see new constraints on the nonlinear parameters, we are allowed to assume any form of the stress tensor as long as it does not contradict our previous assumptions. With this $T_{\mu\nu}$, (\ref{d2Sgen}) simplifies to,
\begin{align}\label{d2ST0}
\Delta^{(2)}S&=\frac{16\pi^2  L^3 R^{10} }{31185 \ \ell_P^3} \left(-\left(\partial _iT_{0 j}\right){}^2\left(-1+60 n_1+336 n_2\right)+6 \partial _iT_{0 j}\partial ^jT_0{}^i\left(-1+n_1\right)\right.\nonumber\\&
\left.+\left(\partial _iT_{j k}\right){}^2\left(-28+60 n_1+168 n_2\right)+ \partial _iT_{k j}\partial ^kT^{i j}\left(5-6 n_1\right)+\left(\partial _iT\right){}^2\left(2+168 n_2\right)\right)\,.
\end{align}
We will use the diagonalization (\ref{diagsumEnst}) once again. For arbitrary $n_1$ and $n_2$, the eigenvalues of $M$ are,
\begin{align}
& \left\{\left(-\frac{2 \left(-7+66 n_1+336 n_2\right)}{31185}\right)^{\times 3},\left(-\frac{2 \left(5+54 n_1+336 n_2\right)}{31185}\right)^{\times 4},\left(-\frac{2 \left(5+54 n_1+336 n_2\right)}{10395}\right),\right.\nonumber\\&
\left(\frac{2 \left(-61+126 n_1+336 n_2\right)}{31185}\right)^{\times 2},\left(\frac{4 \left(-23+54 n_1+168 n_2\right)}{31185}\right),\nonumber\\&
\left(\frac{-79+174 n_1+504 n_2\pm\sqrt{629+3060 n_1{}^2+336 n_2 \left(-23+84 n_2\right)+12 n_1 \left(-227+1512 n_2\right)}}{31185}\right)^{\times 3},\nonumber\\&
\left.\left(\frac{-117+282 n_1+1512 n_2\pm\sqrt{4765+26388 n_1{}^2+1680 n_2 \left(-43+420 n_2\right)+12 n_1 \left(-1867+14952 n_2\right)}}{31185}\right)^{\times 3}\right\}\,. 
\end{align}
For relative entropy to be positive we have to demand each of them to be negative. Moreover, since now there is no $\mathcal{O}(TT)$ term in $\Delta^{(2)}S$, these are completely new constraints on $n_1$ and $n_2$, independent of the previous constraints. Figure \ref{oldnew} shows the new constrained region in the $n_1,n_2$ parameter subspace.

Quite remarkably, the allowed region is a triangular figure very similar to the the one we had obtained from constraining the $\mathcal{O}(TT)$ terms. Since these are independent constraints, the real allowed values of $n_1$ and $n_2$ must lie in the region intersecting the two triangles.

\begin{figure}[!htpb]
\begin{subfigure}{.3\textwidth}
  \centering
  \fbox{\includegraphics[width=\textwidth]{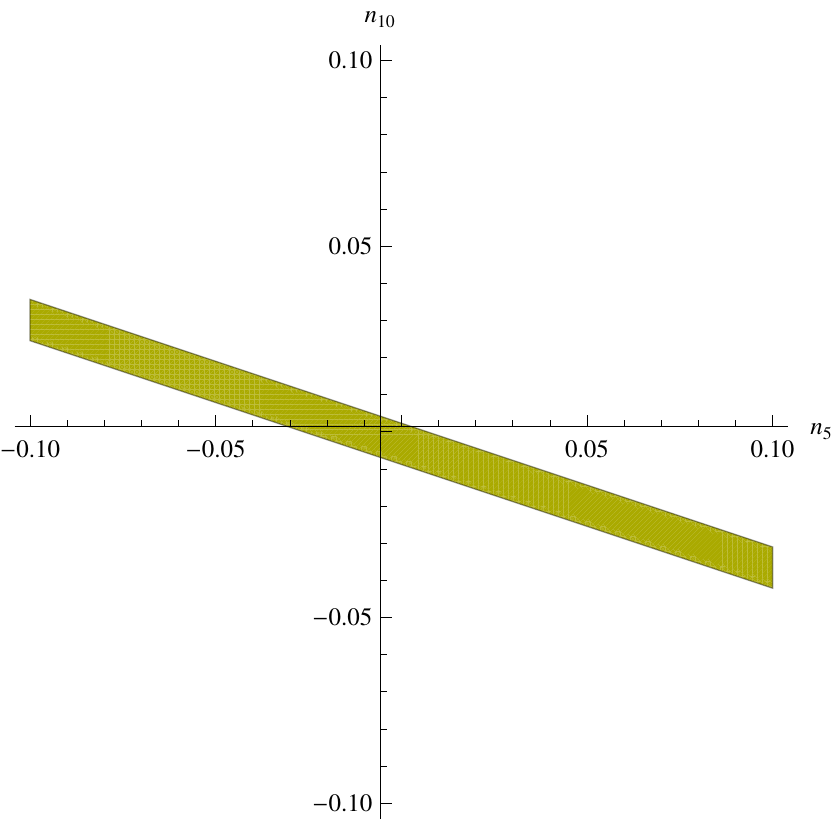}}
  \caption{}
  \label{fig:sub1}
\end{subfigure}\hfill%
\centering
\begin{subfigure}{0.3\textwidth}
  \centering
  \fbox{\includegraphics[width=\textwidth]{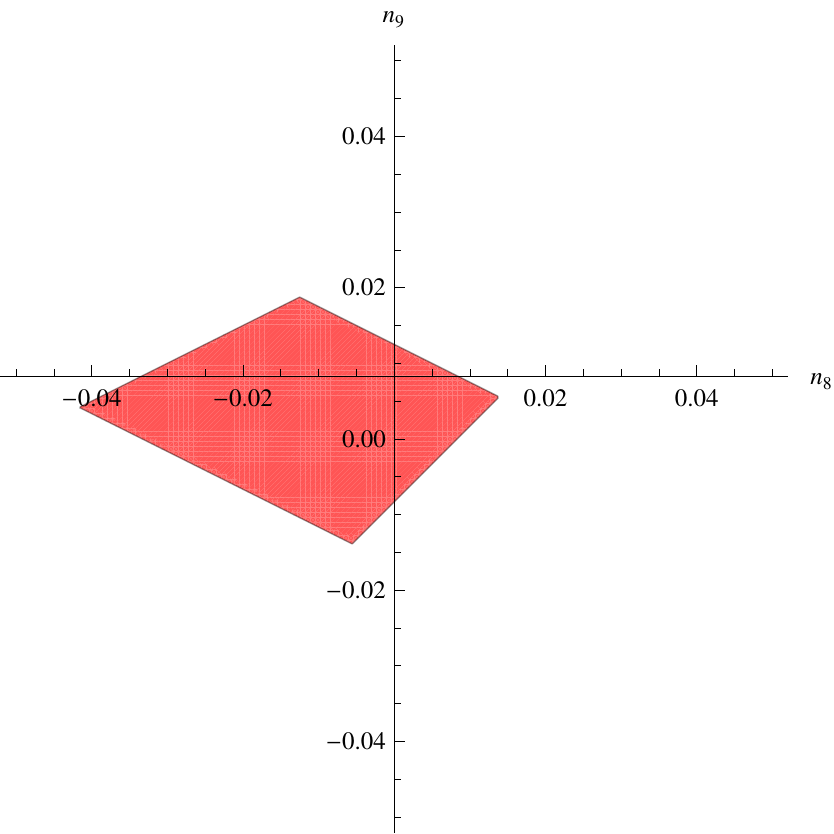}}
  \caption{}
  \label{fig:sub2}
\end{subfigure}\hfill%
\begin{subfigure}{0.3\textwidth}
  \centering
   \fbox{\includegraphics[width=\textwidth]{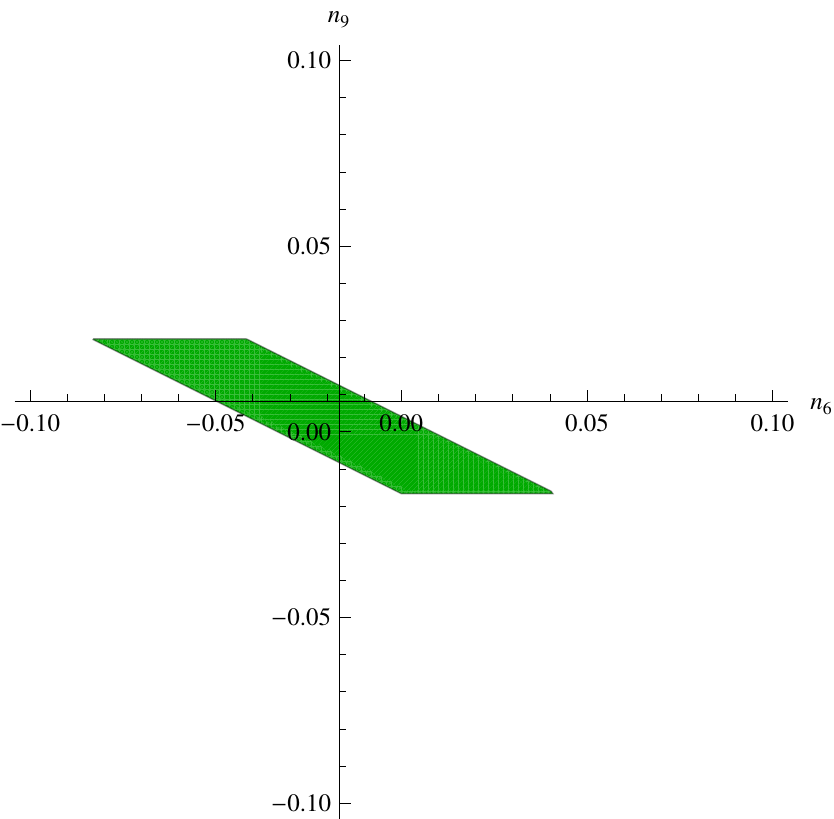}}
  \caption{}\label{n8n9}
  \label{fig:sub3}
\end{subfigure}\hfill
\caption{(a):The plot shows the allowed region for $n_5$ and $n_{10}$ subspace. The region is unbounded in two directions due to lack of enough constraint conditions. The axes origin is chosen to be the Einstein point. (b): The allowed region for ($n_8,n_9$) slice. It is bounded in all directions. (c): The allowed region for ($n_6,n_9$) slice.}
\label{otherns}
\end{figure}

What if we had chosen a different subspace than $(n_1,n_2)$? Say, we wanted to constrain the subspace of $n_3$ and $n_4$. These parameters correspond to $(T_{\mu\alpha}\square T^{\alpha}_\nu+T_{\nu\alpha}\square T^{\alpha}_\mu)$ and $T_{\alpha\beta}\square T^{\alpha\beta}$. Since we had to eliminate the $\mathcal{O}(T\partial\partial T)$ terms in order to check the sign of $\Delta^{(2)}S$, it is not possible to constrain the parameters related to these quantities, such as $n_3$ and $n_4$ with the method discussed above. The same holds for $n_7, n_{12}$ and $n_{13}$. To constrain them we must consider one higher order in the derivative expansion. However it is possible to constrain the rest of the parameters. The above technique does give us some conditions on $n_5, n_6, n_8, n_9, n_{10}$ and $n_{11}$ and they do confine the allowed region in the parameter space (see figure \ref{otherns}). For some of them the 2-dimensional subspaces are bounded to small regions. For others, the conditions are not sufficient to constrain them in all directions. But, it is always possible that we get further constraints on these parameters by looking at higher derivative orders. The important point is we do get bounds in the parameter space from the derivative expansion. It will be interesting to see what new constraints are obtained from higher order corrections.
%

We made a curious observation that at the Einstein values of all parameters, the $\mathcal{O}(T\partial\partial T)$ `cross' terms disappear. We do not have a good answer to how this happens, or what it implies. The cross terms can take both signs. So, till now their absence only serves to make it easy to check the sign of $\Delta^{(2)}S$ at the Einstein point. If they were present, we would have had to look at $\mathcal{O}(\partial\partial T\partial\partial T)$ corrections to the entropy. This is the problem that we face for a general stress tensor, with arbitrary parameters. It would have been wonderful if there was some argument that prevents the cross terms to be present. Then we would have obtained the exact results of nonlinear Einstein equations from entanglement entropy. However it is unlikely that this can come from positivity of relative entropy, since it is an inequality and what we are talking of is a statement of equality. But it does give hope of shedding some light on the bigger question: how are Einstein equations encoded in entanglement entropy? This deserves further investigation.

\section{Constraints in arbitrary dimensions}
\begin{figure}[ht]
\centering
\includegraphics[scale=0.5]{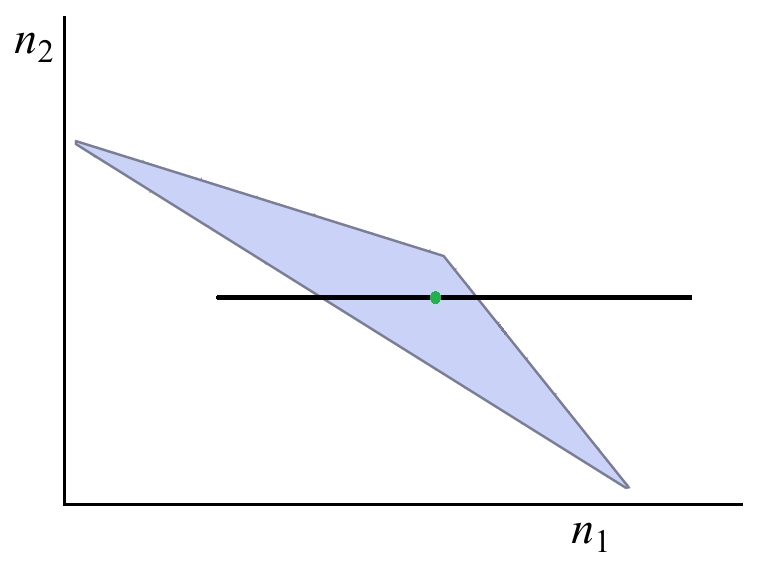}
\caption{The triangle schematically shows the constrained region obtained from the derivatives of the stress tensor in arbitrary dimension. For large dimensions, the triangle is reduced to the line shown in black. It is the same line as obtained in the constant $T_{\mu\nu}$ case. The Einstein point $(\frac{1}{2},-\frac{1}{d-1})$ is shown by the green dot.}\label{arbitd}
\end{figure}

The above analysis can be easily extended to arbitrary dimensions. The first step would be to find $z_1$.  It must satisfy the differential equation (\ref{diffeq}) with the source term,
\begin{align}
J=& -\frac{z_0^{d+1}}{2}\left((d-2)T+(d+2)T_x-\frac{z_0^2}{2(d+2)}\left(d \  \partial^2T+(d+4)\frac{x^ix^j}{R^2}\partial^2T_{ij}\right)\right.\nonumber\\&\left.+x^i\partial_i T+2x^i\partial_0{T}_{0j}+ \frac{1}{R^2}x^ix^jx^k\partial_k T_{ij}\right)\,,
\end{align}
Let us make an educated guess that even in an arbitrary $d$, we will have the same form of $z_1$ as in (\ref{z1}). So without having to worry about the Green's function, we easily find the following solution,
\begin{align}
z_1=-R^2  z_0^{d-1} &\left(\frac{T}{2 (d+1)}+\frac{x^ix^jT_{ij}}{2 (d+1) R^2}+\frac{  x^i\partial_i T}{2 (2+d)}+\frac{  x^ix^jx^k\partial_k T_{ij}}{2 (2+d) R^2}+\frac{ x^ix^j\partial_i\partial_j T}{4 (3+d)}\right.\nonumber\\&\left.+\frac{ x^ix^jx^kx^l\partial_i\partial_jT_{kl}}{4 (3+d) R^2}-\frac{ z_0^2\partial^2 T }{4 (d+2) (3+d)}-\frac{z_0^2 x^ix^j\partial^2T_{ij}}{4 (2+d) (3+d) R^2}\right)\,.
\end{align}
For arbitrary $d$, the metric will be,
\be\label{metricarbitd}
\frac{z^2}{L^2} \ g_{\mu  \nu }=\eta _{\mu  \nu }+z^d\left(T_{\mu  \nu }-\frac{z^2\square T_{\mu  \nu }}{2(d+2)}\right)+z^{2d}\left(n_1T_{\mu  \alpha }T_{\nu }{}^{\alpha }+n_2\eta _{\mu  \nu }T_{\alpha  \beta }T^{\alpha  \beta }+z^2\mathcal{T}_{\mu  \nu }\right)\,,
\ee  
where $\mathcal{T}_{\mu\nu}$ is the same as in (\ref{n3n4}). However the Einstein values of the nonlinear parameters will be different. The first two have the values, $n_1=1/2$ and $n_2=-1/(8(d-1))$. We will be keeping these two arbitrary, however, and only put the values of the other parameters,
\begin{align} \label{n3n4valuesarbitd}
& n_3= -\frac{1}{4(d+2)},n_4= \frac{2 d}{2d(d+2)(d-1)8},\ n_5= -\frac{d}{8(d+2)(d^2-1)},\ n_6=-\frac{1}{2(d+1)(d+2)},\nonumber\\& n_7= \frac{d-2}{8(d+2)(d^2-1)},\ n_8= 0,\ n_9= \frac{1}{4(d+2)(d+1)},\ n_{10}= \frac{1}{8(d+2)(d^2-1)},\ n_{11}= 0,\nonumber \\&n_{12}= -\frac{1}{4(d+1)(d+2)},\ n_{13}= \frac{1}{2(d+1)(d+2)}\,.
\end{align}
With this in hand, we can proceed to compute the entropy correction $\Delta^{(2)}S$. The steps are similar to what is shown in appendix C, but more tedious since $d$ is arbitrary. So we will just mention the result,
\begin{align}
\Delta^{(2)}S&=-\frac{2^{-5-d} d \sqrt{\pi } R^{2+2 d}\Gamma \left[-1+d\right]}{(-1+d) (2+d) \Gamma \left[\frac{5}{2}+d\right]}\Big[ \left\lbrace-(2+d) \left(-2+d^2\right) \left(1+8 (-1+d) n_2\right)\right\rbrace T\partial ^2T\nonumber\\&
+\left(\partial _iT\right)^2\left\lbrace4+d-2 d^2-8 (-1+d) (2+d) \left(-2+d^2\right) n_2\right\rbrace\nonumber\\&
+\partial _iT_{0 j}\partial ^jT_0{}^i\left\lbrace-4 \left(-2+d+d^2\right) \left(-1+n_1\right)\right\rbrace+\left\lbrace 4 \left(-1+4 n_1-8 n_2\right)\right.\nonumber\\&
\left.+ \ d \left(-3+8 n_1+16 n_2+2 d \left(3+2 d-8 n_1-4 d n_1-4 \left(-4+d+d^2\right) n_2\right)\right)\right\rbrace\left(\partial _iT_{j k}\right){}^2\nonumber\\&
\left.+\left\lbrace 2 (2+d) (-1+\left(-1+d^2\right) \left(-1+4 n_1\right)+16 n_2+8 (-2+d) d (1+d) n_2\right)\right\rbrace T_0{}^i\partial ^2T_{0 i} \nonumber\\&
+\left\lbrace 2 (2+d) \left(-1+4 \left(-1+d^2\right) n_1+8 (-1+d) \left(-2+d^2\right) n_2\right)\right\rbrace\left(\partial _iT_{0 j}\right){}^2\nonumber\\&
+\left\lbrace 4 (d-1) \left(-1+d \left(-1+n_1\right)+2 n_1\right)\right\rbrace\partial _iT_{k j}\partial ^kT^{i j}\Big]+ \ \mathcal{O}(TT)
\end{align}
As before, to find constraints on $n_1$ and $n_2$, we have to write the above expression as a sum of squares. For that, we need to set $T_{\mu\nu}(\vec{x}=0)=0$. The matrix $V$, consisting of the independent structures, will be $\frac{1}{2} \left(2-5 d+d^3\right)$-dimensional. Now, if we write $\Delta^{(2)}S$ as $V^T M V$ as before and diagonalize $M$, we get the following eigenvalues,
\begin{align}
& 2b-c,\  2(b+c),\  \frac{1}{2} \left(3 b+c\pm\sqrt{b^2+2 b c+5 c^2}\right),\  e-f,\ e+f,\  4(e+f), \nonumber\\ &\frac{1}{2} \Big( b-c+2 a (-2+d)+(b+c) d  \nonumber\\& \pm\sqrt{4 a^2 (d-2)^2+8 b c (d-2)^2-4 a c (d-1)+b^2 (d-1)^2+c^2 (5+(d-2) d)}\Big)\,.
\end{align}
where,
\begin{align}
&a=\frac{\sqrt{\pi } 2^{-d-5} d \left(8 (d-1) (d+2) (d^2-2) n_2+2 (d+1)^2-5 (d+1)-1\right) \Gamma [d-1]}{(d-1) (d+2) \Gamma \left[d+\frac{5}{2}\right]}\,,\nonumber\\&
b=2^{-5-d} d \left(-1+16 n_2+(1+d) \left(3-16 n_1+40 n_2+2 (1+d) \left(3-2 (1+d)-4d n_1\right.\right.\right.\nonumber\\& \ \ 
+4 (-1+(-2+d) (1+d)) n_2))) \sqrt{\pi } \Gamma[d-1]/(-1+d) (2+d) \Gamma\left[\frac{5}{2}+d\right]\,,\nonumber\\&
c=-\frac{\sqrt{\pi } 2^{-d-3} d \left((d+2)n_1-(d+1)\right) \Gamma [d-1]}{(d+2) \Gamma \left[d+\frac{5}{2}\right]}\,,\nonumber\\&
e=-\frac{\sqrt{\pi } 2^{-d-4} d \Gamma [d-1] (4 (d-1) (d+1) n_1+8 (d-2) d (d+1) n_2+16 n_2-1)}{(d-1) \Gamma \left[d+\frac{5}{2}\right]}\,,\nonumber\\&
f=\frac{\sqrt{\pi } 2^{-d-3} d (n_1-1) \Gamma [d-1]}{\Gamma \left[d+\frac{5}{2}\right]}\,,
\end{align}
with increasing degeneracies for increasing $d$. In order for  $\Delta^{(2)}S$ to be negative, if we demand all these eigenvalues to be negative, we again get a triangular region in the ($n_1$,$n_2$) subspace. With increasing $d$, we observe that the triangle gets thinner and bends towards the $n_2=0$ line. As before, for a certain $d$, this region (Figure \ref{arbitd}) looks very similar to the one obtained from constant stress tensor constraints.  Note that, the last eigenvalue is valid only for odd $d$. For even $d$, it will be different, while we will get even more eigenvalues as $d$ increases. However, they will lead to the same triangular region that will keep getting thinner and bend towards the $n_2=0$ line with increasing $d$, with no extra constraint coming from the new eigenvalues. This has been checked explicitly for $d=4, 6$ and 8.

An interesting lesson that we can learn from the arbitrary $d$ analysis, is what happens to the region for $d\to\infty$. In any dimension, the edges of the triangular region are given by the equation formed by setting the above eigenvalues to zero. The vertices will then be given by solving these equations pairwise. These points don't have  simple expressions, so we will not show them here. But if we take the large $d$ limit of the points, they reduce to $(0,1)$, $(1/2,0)$ and $(1,0)$. So the triangular region becomes a straight line $0<n_1<1$ at $d\to\infty$. Curiously, this is the same line that was obtained for $d\to\infty$ from the constant stress tensor constraints \cite{constrgrav}.

\section{Discussion}

In this paper, we used the positivity of relative entropy to constrain nonlinear perturbations in the metric. We took an $AdS_5$ metric and perturbed the boundary with a space-dependent stress tensor, at both linear and nonlinear level. We demanded that the correction to $\Delta^{(2)}S$ coming from the nonlinear terms, has to be negative to ensure the positivity of relative entropy.  We were motivated mainly by two questions. First, is relative entropy positive for any stress tensor, which is not necessarily constant? Second, what information about gravity can we get from space-dependence of the stress tensor? 

One major finding of this paper was the Green's function, needed to calculate the correction $z_1(x)$ to the Ryu-Takayanagi area functional. It was needed to find the nonlinear correction $\Delta^{(2)}S$ to the entanglement entropy. The correction $z_1(x)$ satisfies a differential equation. This equation had to be solved by guessing, in the original work \cite{myers} for a constant perturbation. Without a proper Green's function, it would not be possible to solve it for more general perturbations. The key  was to observe that the minimal surface deformation was like a propagating massive scalar field on $AdS_3$. Thus, bulk to bulk propagator of the scalar field gave us the Green's function. So, we could develop a systematic way to do analytic computations with nontrivial perturbations to the boundary.

We perturbed the boundary metric with a slowly varying space-dependent stress tensor up to a quadratic order. We neglected contributions coming from more than two derivatives by assuming a derivative expansion. We calculated $z_1$ from the Green's function  and that gave us the correction $\Delta^{(2)}S$ upto two derivatives of $T_{\mu\nu}$. We found that this quantity was manifestly negative, if the nonlinear parameters of the metric assumed values that solve Einstein equations.  Finally we tried to find new constraints on the nonlinear parameters appearing in the metric, by keeping them arbitrary. First we chose to stay only on the subspace of $n_1$ and $n_2$, by fixing the other parameters at the Einstein values. These were the parameters corresponding to zeroth order of the derivative expansion of $T_{\mu\nu}$. We assumed that the stress tensor is zero at the centre of the ball. Then demanding that $\Delta^{(2)}S$ be negative, we obtained new constraint conditions on $n_1$ and $n_2$. These conditions allowed $n_1$ and $n_2$ to be in a bounded region. 

In summary, we can arrive at two main conclusions. First of all, the positivity of relative entropy, for Einstein values of the parameters, implies that for any non-constant stress tensor in the metric (at least upto two orders in a derivative expansion), if Einstein equations are satisfied, then the dual theory will be unitary, as expected. Now we find that the positivity condition constrains the nonlinear parameters in a bounded region. Since the constrainted region found from two derivatives of the perturbation is independent of those found from the constant part,  $n_1$ and $n_2$ can only be in the overlapping region. Points outside this region correspond to non-unitary theories. Thus introducing coordinate dependence the allowed region gets smaller. Our analysis in finite dimensions indicates two possibilities: either the allowed region may shrink to the Einstein point by considering more involved stress tensors, or we get wider class of theories at the nonlinear level which are unitary. In the latter case it will be interesting to study the theories that are at the boundary of the allowed region. However our analysis was done for a spherical entangling region, perturbed by a boundary stress tensor. So, the tantalizing possibility is the former that one might be able to shrink the region to the Einstein point by considering a more sophisticated situation. That will be equivalent to deriving nonlinear Einstein equations from entanglement entropy, and will be a major breakthrough. In fact we observed some interesting things like vanishing of the cross terms of constant and space-dependent part of the stress tensor from $\Delta^{(2)}S$ at Einstein values of the parameters (\ref{d2Enst}). Considering this it is indeed interesting to see what happens at the next order. However such a calculation is more involved to present here. Another interesting direction would be to repeat our calculations for higher derivative gravity. Then we have to replace the Ryu-Takayanagi functional with higher derivative entropy functionals (for recent work see \cite{hdee}). For example, in Gauss-Bonnet gravity, we may have a third axis in the diagram showing how the constraining region varies with the higher derivative parameter. The techniques used in this paper will be useful in answering these questions. 

\vskip 1cm

{\bf Acknowledgements}: The work of SB was supported by World Premier International Research Center Initiative (WPI), MEXT, Japan. AS acknowledges support from a Ramanujan fellowship, Govt. of India.

\appendix

\section{Area functional for constant stress tensor}\label{areafuncconst}
\noindent In this appendix we review the calculation of \cite{constrgrav} to obtain $\Delta^{(2)}S$ for a general constant stress tensor with arbitrary $n_1$ and $n_2$. This will be helpful to understand the more complicated non-constant case. We start with the Ryu-Takayanagi prescription for calculating entanglement entropy in holography,
\be\label{RTapp}
S=\frac{2\pi}{\lp^{d-1}}\int d^{d-1}x\sqrt{h}\,.
\ee
From Taylor expansion one can show that the quadratic correction to $\sqrt{h}$ is,
\be\label{sqrth}
\delta^{(2)}\sqrt{h}=\frac{1}{8}\sqrt{h}(h^{ij}\delta h_{ij})^2+\frac{1}{4}\sqrt{h}\ \delta h^{ij} \delta h_{ij}+\frac{1}{4}\sqrt{h}\ h^{ij}\delta^{(2)} h_{ij}\,.
\ee
The induced metric is given by $h_{ij}=(L^2/z^2)\left( g_{ij}+\partial_i z \partial_j z\right)$. This has to be evaluated at the Ryu-Takayanagi extremal surface. Since we are considering the entanglement of a ball of radius $R$, the extremal surface is given by,
\be\label{z}
z=z_0+z_1=\sqrt{R^2-r^2}+z_1\,.
\ee
Here the part $z_1$ is the result of deformation of the metric given in (\ref{metricpert}). $z_1$ is obtained by plugging (\ref{z}) into (\ref{sqrth}) and then minimizing it. In $\Delta^{(2)}S$, we get 3 kinds of second order contributions,
\vskip 0.5cm
\be\label{3cont}
\Delta^{(2)}S=\frac{2\pi}{\ell_P^{d-1}}\int d^{d-1}x\ \delta^{(2)}\sqrt{h}=\frac{2\pi}{\ell_P^{d-1}}(A_{2,0}+A_{2,1}+A_{2,2})\,.
\ee
This grouping is done according to powers of $z_1$ appearing in the second index. Hence $A_{2,0}$ contains only $\mathcal{O}(TT)$ terms. We compute it by setting $z_1=0$. Then we get from (\ref{sqrth}),
\vskip -0.5cm

\begin{align}\label{A20constT}
A_{2,0}=& L^{d-1}\int d^{d-1}x\ Rz_0^d \left( T_{i0}T^{i0}\left( \frac{n_1}{2}+(d-1)n_2+n_2 \frac{r^2}{R^2} \right)+{(T_{00})}^2\left( \frac{n_2}{2}(d-1)-\frac{n_2r^2}{2R^2} \right)\right.\nonumber\\&+T_{ij}T^{ij}\left( \frac{n_1}{2}+\frac{n_2}{2}(d-1)-\frac{n_2r^2}{2R^2}-\frac{1}{4} \right)-\frac{n_1}{2R^2}x^ix^j T_{i0}T^0_j+x^ix^j T_{ik}T^k_j\left(\frac{1}{2R^2}- \frac{n_1}{2R^2} \right)\nonumber\\&\left.+\frac{1}{8}\left( T^2-T_x^2-2TT_x \right)\right).
\end{align}
\vskip -0.5cm
Here $T_x=x^ix^jT_{ij}/R^2$. The terms $A_{2,1}$ and $A_{2,2}$ are same as in \cite{myers} . We quote their result,
\begin{align}
&A_{2,1}=L^{d-1}\int d^{d-1}x \frac{R}{2z_0}\left[ T\left( z_1-\frac{z_0^2}{R^2}x^i\partial_i z_1 \right)+\frac{T_{ij}}{R^2}\left( 2z_0^2x^i\partial^jz_1-z_1x^ix^j-\frac{z_0^2x^ix^jx^k\partial_kz_1}{R^2} \right) \right]\,, \label{A21constT}\\
&A_{2,2}=L^{d-1}\int d^{d-1}x\frac{R}{z_0^d}\left[ \frac{d(d-1)z_1^2}{2z_0^2}+\frac{z_0^2(\partial z_1)^2}{2R^2}-\frac{z_0^2 (x^i\partial_iz_1)^2}{2R^4}+\frac{(d-1)x^i\partial_iz_1^2}{2R^2} \right]\,.\label{A22constT}
\end{align}
Minimizing $A_{2,1}+A_{2,2}$ w.r.t. $z_1$ gives the differential equation (\ref{gf}). The solution can be found by guessing,
\be\label{z1constTapp}
z_1=-\frac{R^2z_0^{d-1}}{2(d+1)}(T+T_x)\,.
\ee
Plugging this in (\ref{3cont})  and summing we get,
\be\label{7coeff}
\int d^{d-1}x\ \delta^{(2)}\sqrt{h}=L^{d-1}\int d^{d-1}x\ \left( c_1T^2+c_2T_x^2+c_3T_{ij}^2+c_4T_{i0}T^{i0}+c_5\frac{x^ix^jT_{ik}T_j^k}{R^2}+c_6\frac{x^ix^jT_{i0}T_j^0}{R^2}+ c_7TT_x \right)\,,
\ee
\vskip -0.2cm
where the coefficients $c_1\cdots c_7$ are,
\begin{align}
c_1&=\frac{(R^2-r^2)^{(d-4)/2}}{8(1+d)^2R}\left(-4(1+d)^2n_2(r^2-R^2)^2(r^2-(d-1)R^2)\right.\nonumber\\& \ \ \ \ \left.+R^2(2(d^2+2d-1)r^4+(1-5d^2)r^2R^2+(2d^2-d-1)R^4) \right)\,,\\
c_2&=\frac{\left(-r^2+R^2\right)^{\frac{1}{2} (-4+d)} \left(\left(1-5 d^2\right) r^2 R^3+(-3+d (3+4 d)) R^5\right)}{8 (1+d)^2}\,,\\
c_3&=\frac{\left(-r^2+R^2\right)^{d/2} \left(-2 n_2 r^2+(-1+2 n_1+2 (-1+d) n_2) R^2\right)}{4 R}\,,\\
c_4&=\frac{\left(-r^2+R^2\right)^{d/2} \left(n_1 R^2-2 n_2 \left(r^2-(-1+d) R^2\right)\right)}{2 R}\,,\\
c_5&=\frac{\left(d^2-(1+d)^2 n_1\right) R \left(-r^2+R^2\right)^{d/2}}{2 (1+d)^2}\,,\\
c_6&=-\frac{n_1}{2}  R \left(-r^2+R^2\right)^{d/2}\,,\\
c_7&=\frac{(-1+d) R^3 \left(-r^2+R^2\right)^{\frac{1}{2} (-4+d)} \left((1-3 d) r^2+(1+2 d) R^2\right)}{4 (1+d)^2}\,.
\end{align}
Now we integrate the expression (\ref{7coeff}) over the $(d-2)$-sphere on the boundary. We use the trick,
\be\label{trick}
\int d^{(d-1)}x \ f(r)x^ix^jx^kx^l\cdots \mbox{$n$ pairs}=N_n(\delta_{ij}\delta_{kl}\cdots+\mbox{permutations})\int d^{d-1}x \ f(r) r^{2n}\,,
\ee
where $N_n$  some normalization constant. It can be determined by taking the integrand to be $r^{2n}$,
\begin{eqnarray}\label{N123}
N_1&=&\frac{1}{d-1}\hspace{1 cm}\mbox{for $n=1$} \label{N1} \\ 
N_2&=&\frac{1}{\left((d-1)^2+2(d-1)\right)}\hspace{1 cm}\mbox{for $n=2$} \label{N2}\\ 
N_3&=&\frac{1}{\left((d-1)^3+6(d-1)^2+8(d-1)\right)}\hspace{1 cm}\mbox{for $n=3$}\,.\label{N3}
\end{eqnarray} 
This will be used repeatedly for the non-constant case. We use it carry out the integration and obtain the final result (\ref{del2SconstT}).

\section{To find the solution of $z_1$}
\subsection{For a constant stress tensor}\label{const T}
Here we will apply the strategy formulated in section \ref{gensol} to find $z_1$ for a constant stress tensor. In this case, the source is given by,
\begin{equation}\label{const T source}
J = -{z_0}^5 (T+3T_x)
\end{equation}
where $T=T_i^i$, $T_x = x_ix_jT_{ij}/R^2$ and $z_0 = \text{sech}\eta = \frac{1}{X_4}$. 
There are easier ways to find the solution for a constant $T_{\mu\nu}$. But for the non-constant case, it proves to be useful to work in fourier space, because the whole space-dependence of the source then shrinks to just $\exp(i\vec{k}.\vec{x})$. We will follow the same route for the constant case as well. We take the stress tensor to be of the form.
\begin{equation}
T_{ij} = \epsilon_{ij}(\vec k) e^{i\vec k. \vec x}, \ \ \epsilon_{ij}(\vec k)k_j = 0 \,.
\end{equation}
Since the stress tensor is a constant, the only possible mode is. Since the actual source is of the form (\ref{const T source}), $z_1$ will be given by,
\begin{equation}\label{constant T formula}
z_1 = -\left(\epsilon_{ij}\delta^{ij} + 3\frac{\epsilon_{ij}}{R^2}\left(\frac{1}{i}\frac{\partial}{\partial k_i} \frac{1}{i}\frac{\partial}{\partial k_j}\right)\right)_{k=0} \int d^3\hat{x} G(x,\hat{x}) ({z_0}^5 \ e^{i\vec k . \vec{\hat{x} }})\,.
\end{equation}
It is not necessary to evaluate the integral exactly. In fact, since the derivative outside is evaluated at $k=0$ it is sufficient to work out the integral upto two powers of $k$. Let us simplify things a bit by assuming $\vec{k}$ to be along the $x_3$ direction. Then the integral simplifies to 
\be\label{ekz}
\int d^3\hat{x} G(x,\hat{x}) {z_0}^5 \ e^{i\vec k . \vec{ \hat{x}}}= \int d^3\hat{x} G(x,\hat{x}) {(R^2-\hat{r}^2)}^{5/2} \ e^{ik\hat{x}_3}\,.
\ee
As discussed in section \ref{gensol}, we have to transform to a new set of intrinsic coordinates $(\eta',\theta',\phi')$, where the point $\vec{x}=(r,\theta,\phi)=(R \tanh \eta, \theta,\phi)$ becomes the origin. In terms of these coordinates, 
\begin{align}
&\left(1-\frac{r^2}{R^2}\right)=\frac{1}{(\cosh  \eta  \cosh  \eta' +\cos  \theta'  \sinh  \eta  \sinh  \eta' )^2}\\
&\frac{x_3}{R}=\frac{\cos  \theta  \cosh  \eta'  \sinh  \eta +\cos  \theta  \cos  \theta'  \cosh  \eta  \sinh  \eta' -\cos  \phi'  \sin  \theta  \sin  \theta'  \text{sinh$\eta' $}}{\cosh  \eta  \cosh  \eta' +\cos  \theta'  \sinh  \eta  \sinh  \eta' }
\end{align}
Now we can expand the integrand in powers of $k$ and obtain the following integrals\footnote{The integration variables should be written $(\hat{\eta}',\hat{\theta}',\hat{\phi}')$. The hats are removed for tidiness.},
\begin{align}
&\int d\eta ' d\theta ' d\phi ' \frac{\sinh  \eta '(\cosh  \eta '-\sinh  \eta ')^2}{4\pi }\left(R^2-r(\eta ',\theta ',\phi ')^2\right)^{5/2}=\frac{R^5\text{sech}^3\eta}{12 }\,, \label{sol k1}\\
&\int d\eta ' d\theta ' d\phi ' \frac{\sinh  \eta '(\cosh  \eta '-\sinh  \eta ')^2}{4\pi  }i k\  x_3(\eta ',\theta ',\phi ')\left(R^2-r(\eta ',\theta ',\phi ')^2\right)^{5/2}=\frac{i k R^6}{20}\text{  } \cos  \theta  \text{sech}^3 \eta  \tanh  \eta \label{sol k2}\,, \\
&\int d\eta ' d\theta ' d\phi ' \frac{\sinh  \eta '(\cosh  \eta '-\sinh  \eta ')^2}{4\pi  }\left(i k \  x_3(\eta ',\theta ',\phi ')\right){}^2\left(R^2-r(\eta ',\theta ',\phi ')^2\right)^{5/2}\nonumber\\&=\frac{(ik)^2R^7}{360}\text{  }\text{sech}^3\eta  \left(1+6 \cos ^2\theta  \tanh ^2\eta \right) \label{sol k3}\,.
\end{align}
Adding the above the three results we get the r.h.s. of (\ref{ekz}),
\be
\int d^3\hat{x} G(x,\hat{x}) {(R^2-\hat{r}^2)}^{5/2} \ e^{ik\hat{x}_3}=\frac{R^5}{12\cosh ^3\eta }+\frac{ikR^6}{20}\text{  } \cos  \theta  \text{sech}^3 \eta  \tanh  \eta-\frac{k^2R^7}{360}\text{  }\text{sech}^3\eta  \left(1+6 \cos ^2\theta  \tanh ^2\eta \right)\,.
\ee
It is easy to guess the solution for an aribtrary $\vec{k}$. The $\text{sech}^3\eta$ multiplying all the terms is nothing but an overall factor of $(R^2-r^2)^{3/2}=z_0^3$. Every $\cos \theta$ indicates a dot product $\vec{k}.r$. So for general $\vec{k}$,
\be
\int d^3\hat{x} G(x,\hat{x}) {z_0}^5 \ e^{i\vec k . \vec {\hat{x}}}=z_0^3R^2\left(\frac{1}{12}+\frac{i}{20}(k.r)-\frac{\left(k^2R^2+6(k.r)^2\right)}{360}+\cdots\right)\,.
\ee
Now using the formula (\ref{constant T formula}), and replacing $\epsilon_{ij}=T_{ij}$ for $\vec{k}=0$,
\be
z_1=-\frac{z_0^3R^2}{10}\left(T+T_{ij}\frac{x^ix^j}{R^2}\right)\,,
\ee
which is the solution of $z_1$ for constant $T_{\mu\nu}$ in $d=4$ (See \cite{myers}) .

\subsection{For non-constant stress tensor}\label{solnonconstT}
To calculate $z_1$ for non-constant $T_{\mu\nu}$, we need to find the source function $J$. Here the source term will be different from (\ref{const T source}). We can find it by calculating the area functional and minimizing it w.r.t. $z_1$. To begin, we write the area functional as,
\be\label{A2A2012}
\int d^3x \ \sqrt{h}\ =A_2=A_{2,0}+A_{2,1}+A_{2,2}\,.
\ee
It is easy to show that the $A_{2,1}$ in this case is,
\begin{multline}\label{A21}
A_{2,1}=L^{d-1}a\int d^{d-1}x\frac{R}{2z_0}\left( T\left(z_1-\frac{z_0^2}{R^2}x^i\partial_i z_1\right)- \frac{z_0^{2}}{12}\partial^2T\left(3z_1-\frac{z_0^2}{R^2}x^i\partial_i z_1\right)\right.\\\left.+T_{ij}\left(2z_0^2x^i\partial^jz_1/R^2 -\frac{z_1x^ix^j}{R^2}-\frac{z_0^2x^ix^jx^k\partial_kz_1}{R^4}\right) -\frac{z_0^2}{12}\partial^2T_{ij}\left(2z_0^2x^i\partial^jz_1/R^2 -3\frac{z_1x^ix^j}{R^2}-\frac{z_0^2x^ix^jx^k\partial_kz_1}{R^4}\right)\right)\,.
\end{multline}
Note that, since we are calculating entanglement entropy in a time-independent case, all the components of $T_{\mu\nu}$ have been taken to be time-independent.

The $A_{2,2}$ is same as before. $A_{2,0}$ is independent of $z_1$ and it is not required right now. To get the equation for $z_1$, we use,
\be
\frac{\partial{\mathcal{L}}}{\partial z_1}-\partial_i\left(\frac{\partial{\mathcal{L}}}{\partial \partial_i z_1}\right)=0 \hspace{1cm}\text{with}\hspace{1cm}\mathcal{L}=A_{2,1}+A_{2,2}\,.
\ee
The source term comes from $A_{2,1}$. We obtain,
\begin{multline}
\frac{1}{z_0{}^{d-1}R}\left(\partial ^2\left(z_0z_1\right)-\frac{x^ix^j}{R^2}\partial _i\partial _j\left(z_0z_1\right)\right)=\\
\frac{z_0}{2R}\left(T\left(d-2\right)+T_x\left(d+2\right)-\frac{z_0^2}{12}\left(d\partial^2T+(d+4)\frac{x^ix^j}{R^2}\partial^2T_{ij}\right)+x^i\partial_i T+2x^i\partial_0{T}_{0j}+ \frac{1}{R^2}x^ix^jx^k\partial_k T_{ij}\right)\,,
\end{multline}
where we have retained terms only upto two derivatives of $T_{\mu\nu}$. For $d=4$, comparing with (\ref{const T source}),  the source term is,
\be
J=-\frac{z_0^5}{2}\left(2T+6T_x-\frac{z_0^2}{3}\left(\partial^2T+2\frac{x^ix^j}{R^2}\partial^2T_{ij}\right)+x^i\partial_i T+2x^i\partial_0{T}_{0j}+ \frac{1}{R^2}x^ix^jx^k\partial_k T_{ij}\right)\,.
\ee
To find $z_1$, it is useful to work in Fourier space, as before. The solution is then given by,
\begin{align}
z_1&=\left(\epsilon +\frac{3\epsilon _{i j}}{R^2}\left(\frac{1}{i}\frac{\partial }{\partial k_i}\right)\left(\frac{1}{i}\frac{\partial }{\partial k_j}\right)+\frac{\epsilon }{2}i k_i\left(\frac{1}{i}\frac{\partial }{\partial k_i}\right)+\frac{\epsilon _{i j}}{2R^2}i k^m\left(\frac{1}{i}\frac{\partial }{\partial k_m}\right)\left(\frac{1}{i}\frac{\partial }{\partial k_i}\right)\left(\frac{1}{i}\frac{\partial }{\partial k_j}\right)\right.\nonumber\\&\left.-\frac{1}{6}\left(R^2(i k)^2\epsilon +2(i k)^2\epsilon _{i j}\left(\frac{1}{i}\frac{\partial }{\partial k_i}\right)\left(\frac{1}{i}\frac{\partial }{\partial k_j}\right)\right)+\frac{1}{6}\left((i k)^2\delta _{i j}\left(\frac{1}{i}\frac{\partial }{\partial k_i}\right)\left(\frac{1}{i}\frac{\partial }{\partial k_j}\right)\epsilon\right.\right.
\nonumber\\& \left.\left. +2\frac{(i k)^2}{R^2}\epsilon _{i j}\delta _{a b}\left(\frac{1}{i}\frac{\partial }{\partial k_a}\right)\left(\frac{1}{i}\frac{\partial }{\partial k_a}\right)\left(\frac{1}{i}\frac{\partial }{\partial k_i}\right)\left(\frac{1}{i}\frac{\partial }{\partial k_j}\right)\right)\right) \int d^3\hat{x} G(x,\hat{x}) ({z_0}^5 \ e^{i\vec k . \vec{\hat{x} }})\,.
\end{align}
We again start with a $\vec{k}$ along the $x_3$ direction. But, this time we cannot set $\vec{k}=0$ and (\ref{ekz}) must be expanded upto 4 powers of $k$. We already evaluated the first 3 (eq. (\ref{sol k1}),(\ref{sol k2}),(\ref{sol k3})). The third and fourth powers give,
\begin{align}
\int d\eta ' d\theta ' d\phi ' &\frac{\sinh  \eta '(\cosh  \eta '-\sinh  \eta ')^2}{4\pi  }\left(i k \ x_3(\eta ',\theta ',\phi ')\right){}^3\left(R^2-r(\eta ',\theta ',\phi ')^2\right)^{5/2}= \nonumber\\&-\frac{(i k)^3 R^8\cos  \theta  \text{sech}^3 \eta  \tanh  \eta  \left(3+10 \cos ^2 \theta  \tanh ^2\eta \right)}{2520}\,,\\
\int d\eta ' d\theta ' d\phi ' &\frac{\sinh  \eta '(\cosh  \eta '-\sinh  \eta ')^2}{4\pi  }\left(i k \  x_3(\eta ',\theta ',\phi ')\right){}^4\left(R^2-r(\eta ',\theta ',\phi ')^2\right)^{5/2}=\nonumber\\& \frac{(i k)^4 R^9\text{sech}^3\alpha  \left(1+6 \cos ^2\beta  \tanh ^2\alpha +15 \cos ^4\beta  \tanh ^4\alpha \right)}{20160}\,.
\end{align}
This time too, it is easy to generalize the above for an arbitrary  $\vec{k}$. It is straightforward to evaluate the solution for $z_1$ in Fourier space,
\be
z_1=-z_0{}^3R^2\left(\frac{1}{10}+\frac{i k.r}{12}-\frac{1}{28} (k.r)^2-\frac{k^2 r^2}{168}+\frac{k^2 R^2}{168}\right)(T +T_x)\,.
\ee
Here, $T+T_x$ is evaluated with the value of $T_{ij}$ at the origin. This gives a simple solution in coordinate space given by eq. (\ref{z1}).

\section{Evaluating the area functional for non-constant stress tensor} \label{areafunc}

Here we use the $z_1$ solution (\ref{z1}) to evaluate the area functional $A_2$ (\ref{A2A2012}) step-by-step. The best way to do this is to work out the expressions $A_{2,0}$, $A_{2,1}$ and $A_{2,2}$ separately. Let us begin with $A_{2,1}$. Take eq. (\ref{A21}) and Taylor expand all $T_{ij}$-s around origin.
\begin{align}
& A_{2,1}=
L^3\int d^3x \frac{R}{2z_0}\left(\left(T+x^i\partial _iT+\frac{1}{2}x^ix^j\partial _i\partial _jT\right) z_1-\left(T+ x^i\partial _iT+\frac{1}{2}x^ix^j\partial _i\partial _jT\right)\frac{z_0{}^2}{R^2}x^i\partial _iz_1\right.\nonumber\\&-\frac{z_0{}^2}{12}\partial ^2T\left(3z_1-\frac{z_0{}^2}{R^2}x^i\partial _iz_1\right)
-\frac{x^ix^jT_{i j} z_1}{R^2}-\frac{\text{z0}^2x^ix^jT_{i j} x^k\partial _kz_1}{R^4}+\frac{2z_0{}^2T_{i j}x^i\partial ^jz_1}{R^2}- \frac{z_1}{R^2}x^ix^jx^k\partial _kT_{i j}\nonumber\\&-\frac{z_0{}^2}{R^4}x^i\partial _iz_1 x^kx^lx^j\partial _jT_{k l}
-\frac{1}{2R^2}z_1 x^ix^jx^kx^l\partial _i\partial _jT_{k l}-\frac{1}{2R^4}z_0{}^2x^m\partial _mz_1 x^ix^jx^kx^l\partial _i\partial _jT_{k l}+2\frac{z_0{}^2}{R^2}x^i\partial _iT_{j k}x^j\partial ^kz_1\nonumber\\&
\left.+\frac{z_0{}^2}{R^2}x^ix^j\partial _i\partial _jT_{k l}x^k\partial ^lz_1+z_0{}^2\left(\frac{z_1x^ix^j\partial ^2T_{i j}}{4R^2}+\frac{z_0{}^2x^ix^j\partial ^2T_{i j} x^k\partial _kz_1}{12R^4}-\partial ^2T_{i j}x^i\partial ^jz_1 \frac{z_0{}^2}{6R^2}\right)\right)\,.
\end{align}
Now we put in the solution (\ref{z1}) and simplify using the trick (\ref{trick}). We get,
\begin{align}
& A_{2,1}=\mathcal{O}(TT)+L^3\int \frac{ d^3x}{88200 R^3}(r-R) (r+R) \left(420 \left(\partial _iT_{j k}\right){}^2 r^8+840  r^8\partial _iT_{k j}\partial ^kT^{i j}-1680 \left(\partial _iT_{j k}\right){}^2 r^6 R^2\right.\nonumber\\&
-3360r^6 R^2 \partial _iT_{k j}\partial ^kT^{i j}+980 r^4 R^4\left(\partial _iT_{j k}\right){}^2 +1960  r^4 R^4\partial _iT_{k j}\partial ^kT^{i j}+70 \left(\partial _iT\right){}^2 \left(3 r^8+30 r^6 R^2+35 r^4 R^4\right)\nonumber\\&
+T^{i j}\partial ^2T_{i j}\left(910 r^8-4040 r^6 R^2+3682 r^4 R^4-840 r^2 R^6\right)+455 r^8T\partial ^2T  +3370 r^6 R^2 T\partial ^2T +1246r^4 R^4 T\partial ^2T \nonumber\\&
\left. +1190 r^2 R^6T\partial ^2T  -1365 R^8T\partial ^2T  +12 r^2 \left(35 r^6+116 r^4 R^2-322 r^2 R^4+175 R^6\right) T^{i j}\partial _i\partial _jT\right)
\end{align}
Carrying out the integral,
\be
A_{2,1}=\mathcal{O}(TT)-\frac{8\pi L^3R^{10} \left(283 \left(\partial _iT\right){}^2+7 \left(\partial _iT_{j k}\right){}^2+14 \partial _iT_{k j}\partial ^kT^{i j}+72 T^{i j}\partial _i\partial _jT\right)}{405405}
\ee

Now let us evaluate $A_{2,2}$, given by (\ref{A22constT}),
\be
A_{2,2}=L^3\int d^3x \frac{R}{z_0^4}\left(\frac{6z_1{}^2}{z_0{}^2}+\frac{z_0{}^2\left(\partial _iz_1\right)}{2R^2}-\frac{z_0{}^2}{2R^4}\left(x^i\partial _iz_1\right){}^2+\frac{(d-1)}{R^2}z_1 x^i\partial _iz_1\right)\,.
\ee
Putting the solution (\ref{z1}) and once again using the trick (\ref{trick}) we get,  after integration,
\be
A_{2,2}=\mathcal{O}(TT)+\frac{4\pi L^3R^{10} \left(283 \left(\partial _iT\right){}^2+7 \left(\partial _iT_{j k}\right){}^2+14 \partial _iT_{k j}\partial ^kT^{i j}+72 T^{i j}\partial _i\partial _jT\right)}{405405}
\ee

Finally we come to $A_{2,0}$. The expression for $A_{2,0}$ given in (\ref{A20constT}) does not have any derivative of $T_{\mu\nu}$ since there was no derivative in the metric previously. Now, we will have additional terms due to these derivatives appearing in (\ref{metric}). So let us write $A_{2,0}$ as,
\be
A_{2,0}=A^0_{2,0}+A^1_{2,0}\,,
\ee
where $A^0_{2,0}$ is nothing but our old formula for $A_{2,0}$. We can taylor expand all $T_{\mu\nu}$ around the origin and use (\ref{trick}) to find,
\begin{align}\label{A200}
& A^0_{2,0}=L^3\int d^3x R z_0{}^4\left(\left(-N_1 r^2T_0{}^i\partial ^2T_{0 i}-N_1 r^2\left(\partial _iT_{0 j}\right){}^2\right)\left(\frac{n_1}{2}+(d-1)n_2-n_2 \frac{r^2}{R^2}\right)\right.\nonumber\\&
+N_1n_2 r^2\left(T\partial ^2T+\left(\partial _iT\right){}^2\right)\left(\frac{d-1}{2}- \frac{r^2}{2R^2}\right)+\left(N_1 r^2\left(\partial _iT_{j k}\right){}^2+T^{i j}\partial ^2T_{i j}\right)\left(\frac{n_1}{2}+\frac{n_2}{2}(d-1)-n_2 \frac{r^2}{2R^2}-\frac{1}{4}\right)\nonumber\\&
+\frac{n_1N_2 r^4}{2R^2}\left(T_0{}^i\partial ^2T_{0 i}+\left(\partial _iT_{0 j}\right){}^2+ \partial _iT_{0 j}\partial ^jT_0{}^i\right)+N_2 r^4\left(\frac{1}{2R^2}-\frac{n_1}{2R^2}\right)\left(T^{i j}\partial ^2T_{i j}+ \left(\partial _iT_{j k}\right){}^2+\partial _iT_{k j}\partial ^kT^{i j}\right)\nonumber\\&
+\frac{1}{8}\left( N_1 r^2\left(\left(\partial _iT\right){}^2+T\partial ^2T\right)- N_3 \frac{r^6}{R^4}\left(2\left(\partial _iT_{j k}\right){}^2+\left(\partial _iT\right){}^2+4\partial _iT_{k j}\partial ^kT^{i j}+T\partial ^2T+2T^{i j}\partial _i\partial _jT+2T^{i j}\partial ^2T_{i j}\right)\right.\nonumber\\&
\left.\left.-2N_2 \frac{r^4}{R^2}\left(T\partial ^2T+\left(\partial _iT\right){}^2+T^{i j}\partial _i\partial _jT\right)\right)\right)
\end{align}
To evaluate the other part $A^1_{2,0}$ we go back to (\ref{sqrth}), and keep only the tensor structures containing two derivatives. Then under the integral sign we get, 
\begin{eqnarray}
\left(h_{i j}\text{$\delta h$}^{i j}\right){}^2&=&-\frac{1}{6} z_0{}^{10}\left(T \partial ^2T-N_1 T\partial ^2T\frac{r^2}{R^2}-\frac{r^2}{R^2}N_1 T \partial ^2T+\frac{N_2 r^4}{R^4}\left(T \partial ^2T+2T^{i j}\partial ^2T_{i j}\right)\right)\,,\\
\text{$\delta h$}^{i j}\text{$\delta h$}_{i j} &=& -\frac{1}{6}z_0{}^{10}\left(T^{i j}\partial ^2T_{i j}-\frac{2 N_1}{R^2}r^2T^{i j}\partial ^2T_{i j}+\frac{N_2}{R^4}r^4\left(T \partial ^2T+2T^{i j}\partial ^2T_{i j}\right)\right)\,,\\
h^{ij}\delta^2h_{ij}&=& z_0^{10}\left(\eta^{ij}-\frac{x^ix^j}{R^2}\right)\mathcal{T}_{ij}\,.
\end{eqnarray}
Then using (\ref{n3n4}),
\begin{align}\label{A201}
& A^1_{2,0}=L^3\int d^3x\frac{R \left(-r^2+R^2\right)^3 }{48 }\left(T\partial ^2T\left(-1+\frac{2\text{  }N_1 r^2}{R^2}\right)-\frac{N_2 r^4 \left(2 T^{i j}\partial ^2T_{i j}+T\partial ^2T\right)}{R^4}\right.\nonumber\\&
+2 \left(T^{i j}\partial ^2T_{i j}\left(1-\frac{2N_1 r^2}{R^2}\right)+\frac{N_2 r^4 \left(2 T^{i j}\partial ^2T_{i j}+T\partial ^2T\right)}{R^4}\right)+24 \left(\left(\left(\partial _iT\right){}^2-2 \left(\partial _iT_{0 j}\right){}^2+\left(\partial _iT_{j k}\right){}^2\right) \right.\nonumber\\&
n_{10} \left(3-\frac{r^2}{R^2}\right)+\left(-\partial _iT_{0 j}\partial ^jT^{0 i}+\partial _iT_{k j}\partial ^kT^{i j}\right) n_{11} \left(3-\frac{r^2}{R^2}\right)+2 \left(-T^{0 i}\partial ^2T_{0 i}+T^{i j}\partial ^2T_{i j}\right) n_3 \left(1-\frac{N_1 r^2}{R^2}\right)+\nonumber\\&
\left(\left(\partial _iT\right){}^2-2 \left(\partial _iT_{0 j}\right){}^2+\left(\partial _iT_{j k}\right){}^2\right) n_5 \left(1-\frac{N_1 r^2}{R^2}\right)+\partial _iT_{k j}\partial ^kT^{i j} n_6 \left(1-\frac{N_1 r^2}{R^2}\right)+n_{13} \left(1-\frac{N_1 r^2}{R^2}\right) T^{i j}\partial _i\partial _jT\nonumber\\&
+\left(-\left(\partial _iT_{0 j}\right){}^2+\left(\partial _iT_{j k}\right){}^2\right) n_8 \left(1-\frac{N_1 r^2}{R^2}\right)+2 \left(-\partial _iT_{0 j}\partial ^jT^{0 i}+\partial _iT_{k j}\partial ^kT^{i j} \right) n_9 \left(1-\frac{N_1 r^2}{R^2}\right)+n_4\left(3-\frac{r^2}{R^2}\right)\nonumber\\&
\left.\left. \left(-2T^{0 i}\partial ^2T_{0 i}+T^{i j}\partial ^2T_{i j}+T\partial ^2T\right)+n_7 \left(1-\frac{N_1 r^2}{R^2}\right) \left(-2T^{0 i}\partial ^2T_{0 i}+T^{i j}\partial ^2T_{i j}+T\partial ^2T\right)\right)\right)\,.
\end{align}
Now adding (\ref{A200}) and (\ref{A201}) and integrating, we obtain,
\begin{align}
& A_{2,0}=\frac{16\pi  L^3R^{10}}{135135} \left(12 \partial _iT_{k j}\partial ^kT^{i j}+\left(\partial _iT_{j k}\right){}^2\left(-59+130 n_1\right)+\left(\partial _iT\right){}^2 \left(29+2340 n_{10}+364 n_2+780 n_5\right)\right.\nonumber\\&
-26 T^{0 i}\partial ^2T_{0 i} \left(5 n_1+4 \left(7 n_2+15 \left(n_3+3 n_4+n_7\right)\right)\right)+13 \partial _iT_{k j}\partial ^kT^{i j}\left(-n_1+60 \left(3 n_{11}+n_6+2 n_9\right)\right)\nonumber\\&
+13 \left(-10\text{  }\left(\partial _iT_{0 j}\right){}^2n_1+2 T^{i j}\partial ^2T_{i j} \left(5 n_1+14 n_2+30 \left(2 n_3+3 n_4+n_7\right)\right)+\partial _iT_{0 j}\partial ^jT^{0 i} \left(n_1-60 \left(3 n_{11}+2 n_9\right)\right)\right)\nonumber\\&
\left.\left.-\left(\partial _iT_{0 j}\right){}^2 \left(90 n_{10}+14 n_2+15 \left(2 n_5+n_8\right)\right)+T\partial ^2T \left(7 n_2+15 \left(3 n_4+n_7\right)\right) \right)+\left(-7+780 n_{13}\right)T^{i j}\partial _i\partial _jT\right)\nonumber\\&
+52 \left(\left(\partial _iT_{j k}\right){}^2 \left(45 n_{10}+7 n_2+15 \left(n_5+n_8\right)\right)\right.+\mathcal{O}(TT)\,.
\end{align}
Now we can add the three components, $A_{2,0}$, $A_{2,1}$ and $A_{2,2}$ to arrive at the area functional given by eq. (\ref{A2}).

\end{document}